\documentclass[pra,twocolumn,showpacs,preprintnumbers,amsmath,amssymb,floatfix,aps]{revtex4-1}

\usepackage{dcolumn}
\usepackage{bbm, bm, epsfig, amsmath, amsfonts, amssymb, wasysym, graphicx, color}

\newcommand{\ef}{\varepsilon_{\rm F}}
\newcommand{\kf}{k_{\rm F}}
\newcommand{\betamu}{y}

\begin{document}

\title{Numerical study of the unitary Fermi gas across the superfluid transition}

\author{Olga Goulko}
\affiliation{Department of Physics, University of Massachusetts, Amherst, MA 01003, USA}

\author{Matthew Wingate}
\affiliation{Department of Applied Mathematics and Theoretical Physics, University of Cambridge, Centre for Mathematical Sciences, Cambridge CB3 0WA, United Kingdom}

\begin{abstract}
We present results from Monte Carlo calculations investigating the properties
of the homogeneous, spin-balanced unitary Fermi gas in three dimensions. The
temperature is varied across the superfluid transition allowing us to
determine the temperature dependence of the chemical potential, the energy per
particle and the contact density. Numerical artifacts due to finite volume and
discretization are systematically studied, estimated, and reduced.
\end{abstract}

\maketitle

\section{Introduction}

Ultracold, dilute gases of fermionic atoms have been studied extensively
lately, in part due to the system being the simplest environment with strong
interactions between fermions (for recent reviews of this active field see
e.g.\ Refs.~\cite{review1,zwerger}).  Most remarkably, a three-dimensional
(3d) atomic gas with two hyperfine states, say lithium or potassium, can be
constructed to have resonant interactions: by applying an external
magnetic field the S-wave scattering length $a$ can be tuned to satisfy $1/a =
0$.  In this special situation, the properties of the gas become universal,
dependent only on the density and temperature.  What can be learned in the
atomic physics laboratory then has implications for nonrelativistic fermions
as small as nucleons.

This resonant Fermi gas -- often called the ``unitary Fermi gas'' due to the
scattering being limited only by unitarity -- provides an excellent opportunity
for quantitative theoretical calculations.  The clean separation of scales
means that much can be inferred from dimensional analysis and scaling
arguments.  What remains to be determined are universal dimensionless 
constants, such as the Bertsch parameter \cite{Bishop:2001rf},
or functions of the product of the inverse temperature $\beta$ and the chemical potential $\mu$ \cite{Ho:PRL92}
which completely specify the thermodynamic and hydrodynamic
behavior of the unitary Fermi gas.  Much effort has gone into using 
first-principles numerical methods to determine these quantities (see
Refs.~\cite{lodereview,Lee:2008fa,Drut:2012md} for reviews).  

In this paper we use lattice Monte Carlo methods to give numerical results
for thermodynamic quantities as the temperature is varied through the
superfluid phase transition.  In particular we determine
the chemical potential, mean energy density, and contact density
as functions of temperature.  The corresponding universal functions
$f(\beta\mu)$ are made dimensionless by taking ratios with the appropriate powers of the
Fermi energy $\varepsilon_F$. It is notable that several other
numerical methods for studying the Fermi gas cannot study the superfluid 
phase.  With this study, we also  pay particular attention to investigating
and quantifying the systematic uncertainties associated with taking the
thermodynamic limit \textit{and} the continuum limit, which is crucial to obtain correct physical results \cite{fermihubbardcontlim1,fermihubbardcontlim2}. Generally, we find good agreement with experiment.

In previous work \cite{ourmain} we used the Diagrammatic Determinant Monte Carlo (DDMC) algorithm \cite{rubtsov, burovski, burovskieos} to numerically determine the critical temperature $T_c$ and thermodynamic properties of the unitary Fermi gas at $T=T_c$. Here we study the temperature dependence of physical observables in the approximate range $T_c /2\le T \le 2 T_c$. An approach which is formulated in the continuum, bold-line diagrammatic Monte Carlo (bold DMC), has been used to compute quantities above the critical temperature \cite{vanhouckewerneretal, BDMCvanhoucke2013}, and these results agree well with experimental measurements. This method as presently formulated does not extend to temperatures below $T_c$ due to the singularity in the pair propagator appearing in the superfluid phase. Temperature effects have also been studied in a lattice computation using a hybrid Monte Carlo approach \cite{Drut:2011tf}; however results using different lattice spacings are not presented there.

\section{Setup}
We consider a system of equal-mass fermions with two spin components labeled by the spin index $\sigma=\{\uparrow,\downarrow\}$. Since the details of the physical potential governing the interatomic interactions are irrelevant in the dilute limit realized in cold-atom experiments, we can work on a spatial lattice provided that we also take the dilute limit \cite{chenkaplan}. The Hamiltonian is that of the simple Fermi-Hubbard model in the grand canonical ensemble,
\begin{equation}
H=\sum_{\mathbf{k},\sigma}(\epsilon_\mathbf{k}-\mu_\sigma)c^\dagger_{\mathbf{k}\sigma}c_{\mathbf{k}\sigma}+U\sum_{\mathbf{x}} c^\dagger_{\mathbf{x}\uparrow}c_{\mathbf{x}\uparrow}c^\dagger_{\mathbf{x}\downarrow}c_{\mathbf{x}\downarrow},
\end{equation}
where the first term corresponds to the kinetic part of the Hamiltonian
$H_{\rm kin}$ and the second term to the interaction part $H_{\rm int}$. The
units are chosen such that $\hbar=k_B=2m=1$. We work on a 3d periodic lattice
with lattice spacing $b$ and $L^3$ sites. The discrete dispersion relation
reads $\epsilon_\mathbf{k}=2b^{-2}\sum_{j=1}^{3}(1-\cos{k_jb})$;
$\mu_\sigma$ is the chemical potential and $c^\dagger_{\mathbf{k}\sigma}$
the fermionic creation operator. The coupling constant $U<0$ corresponding to
attractive interaction can be tuned so that the scattering length becomes
infinite. The corresponding value in the infinite-volume limit is $U=-7.914/b^2$
which is the value we use throughout the calculation. Another approach is to
include finite-volume effects in this two-body matching calculation
\cite{Lee:2005is,lee,Drut:2011tf}. It remains to be seen which approach leads
to a milder extrapolation of many-body results to the continuum limit.

The partition function $Z = \mathrm{Tr}\exp(-\beta H)$ can be written as a series of products of two matrix determinants built of free finite-temperature Green's functions \cite{rubtsov}. If $\mu_\uparrow=\mu_\downarrow\equiv\mu$, as is always assumed to be the case in the present work, the two determinants are identical since the spin-dependence enters only via the chemical potential. Consequently all terms in the series are positive, and the series can be used as a probability distribution for Monte Carlo sampling.

We use the DDMC algorithm as introduced in \cite{burovski} with several modifications which increase the efficiency by reducing autocorrelation effects as compared to the original setup. We account for remaining autocorrelations by binning the data to the point where the statistical error is insensitive to the bin size. A detailed description of our numerical setup is given in \cite{goulko_lat2009, ourmain, mythesis}.

\begin{figure}
\centering
\includegraphics[width=\columnwidth]{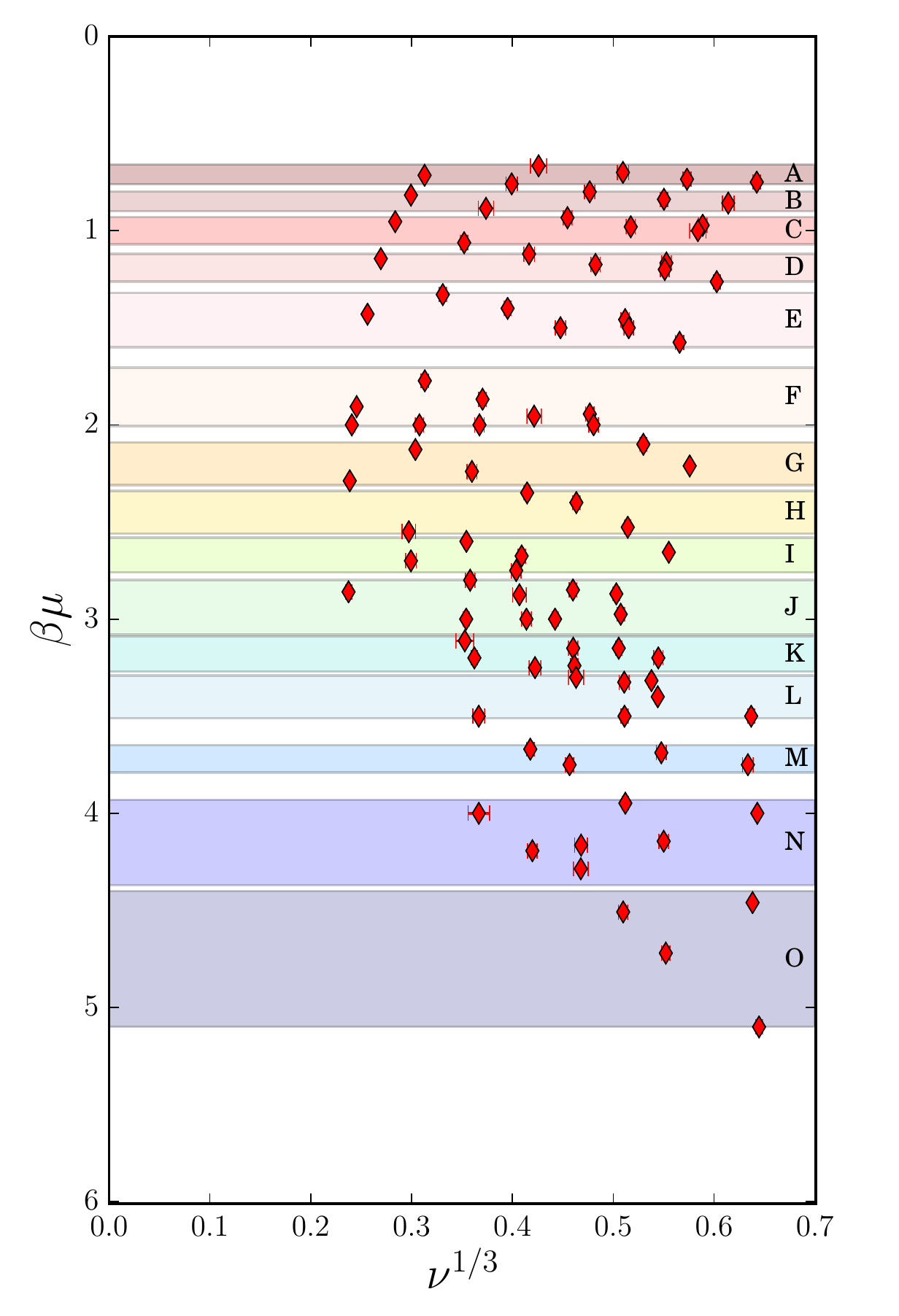}
\caption{\label{fig:betamu_nu13}Diamonds correspond to parameters used for 
individual simulations extrapolated to the thermodynamic limit. 
The 15 bands indicate the subsets of data included 
in the continuum ($\nu\to 0$) extrapolations as described in 
Sec.~\ref{sec:limits}.}
\end{figure}

We performed many calculations, varying the dimensionless inputs for the
  chemical potential ($\mu b^2$) and inverse temperature ($\beta/b^2$), as
  well as the number of lattice points $L^3$, so that controlled
  extrapolations to the thermodynamic and continuum limits could be taken for a
  range of temperatures in both the normal and superfluid phases. Each diamond
  in Fig.~\ref{fig:betamu_nu13} represents the thermodynamic limit of lattice
  calculations done for values of $\mu b^2$ and $\beta/b^2$ resulting in the
  corresponding filling factor $\nu$ and dimensionless product $\beta\mu$.

\section{Thermodynamic and continuum limits}
\label{sec:limits}

\begin{figure*}
\centering
\hspace{-5mm}\includegraphics[width=0.35\textwidth]{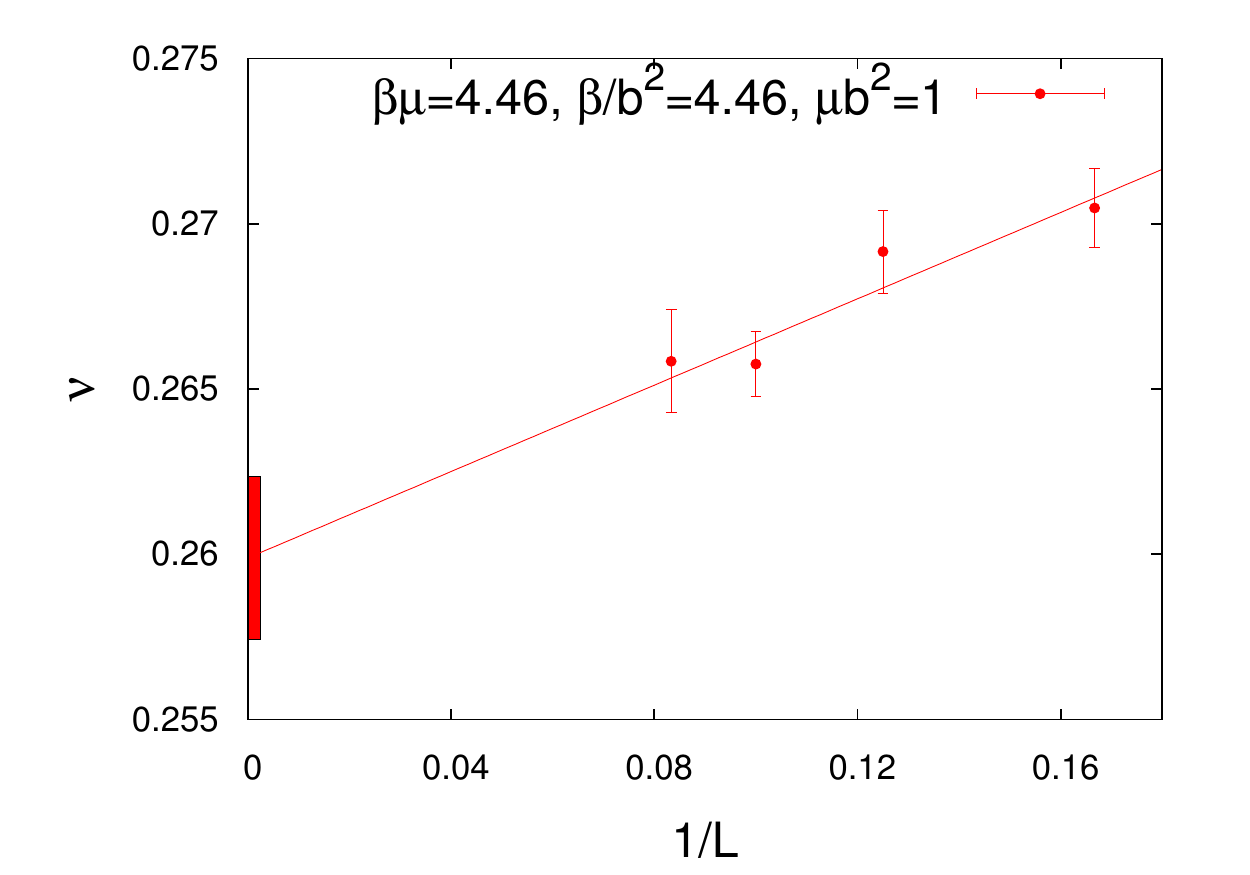}
\hspace{-4mm}\includegraphics[width=0.35\textwidth]{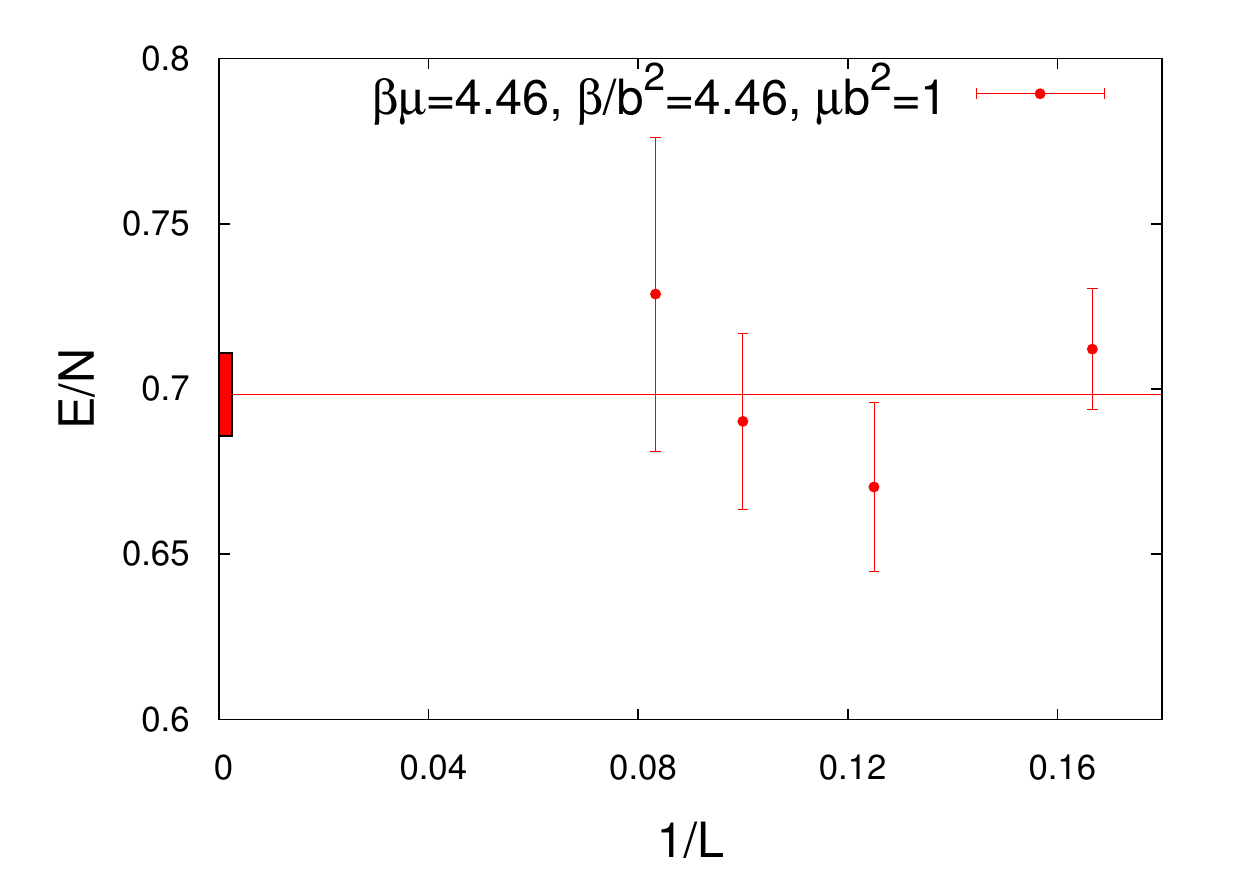}
\hspace{-4mm}\includegraphics[width=0.35\textwidth]{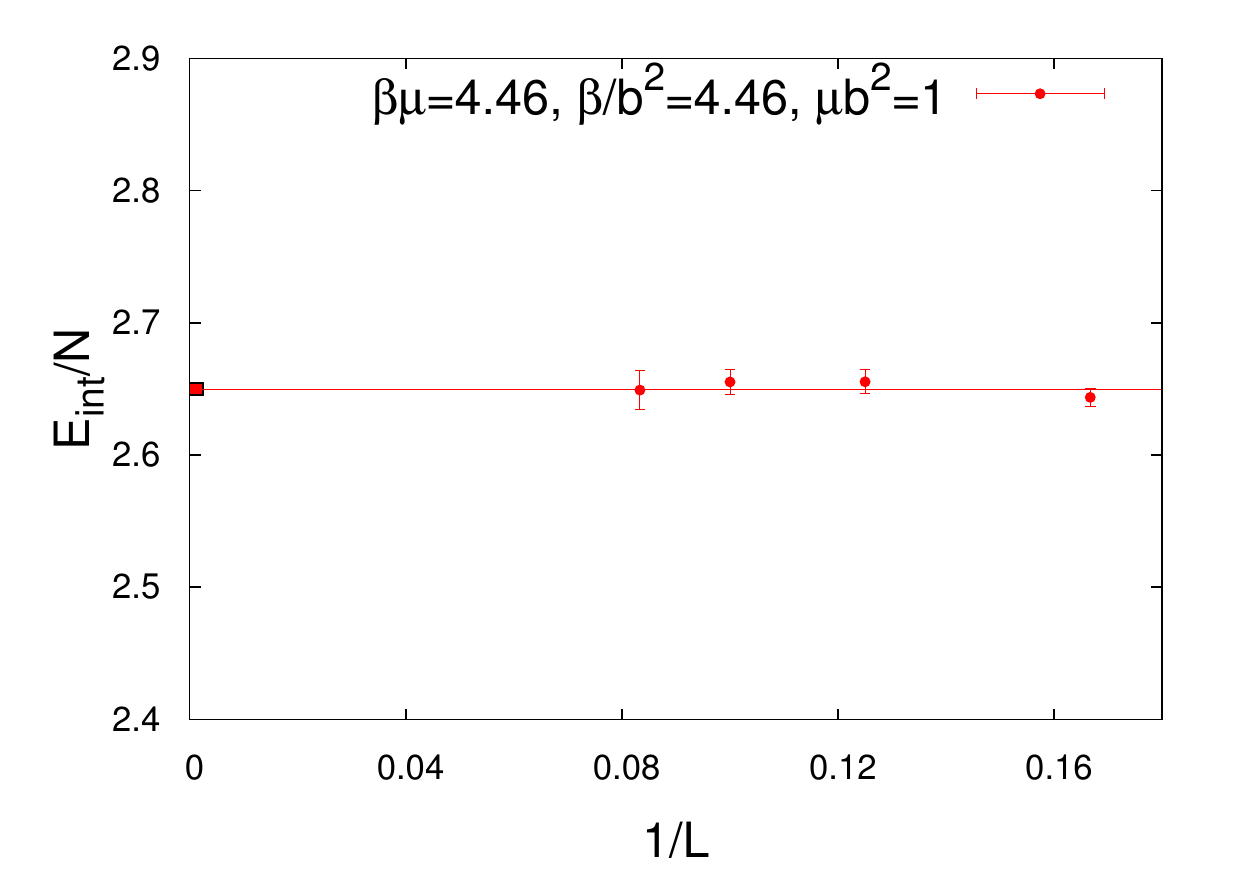}

\hspace{-5mm}\includegraphics[width=0.35\textwidth]{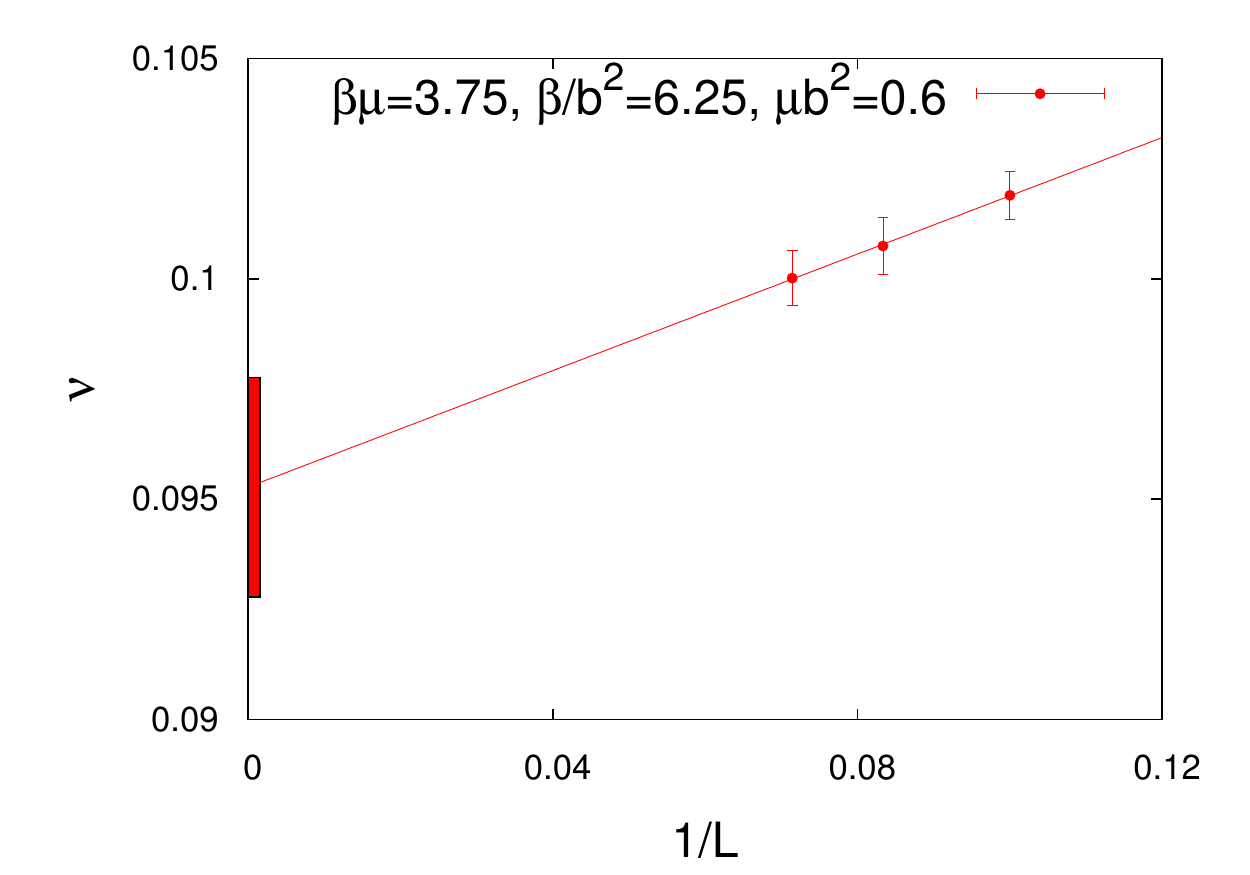}
\hspace{-4mm}\includegraphics[width=0.35\textwidth]{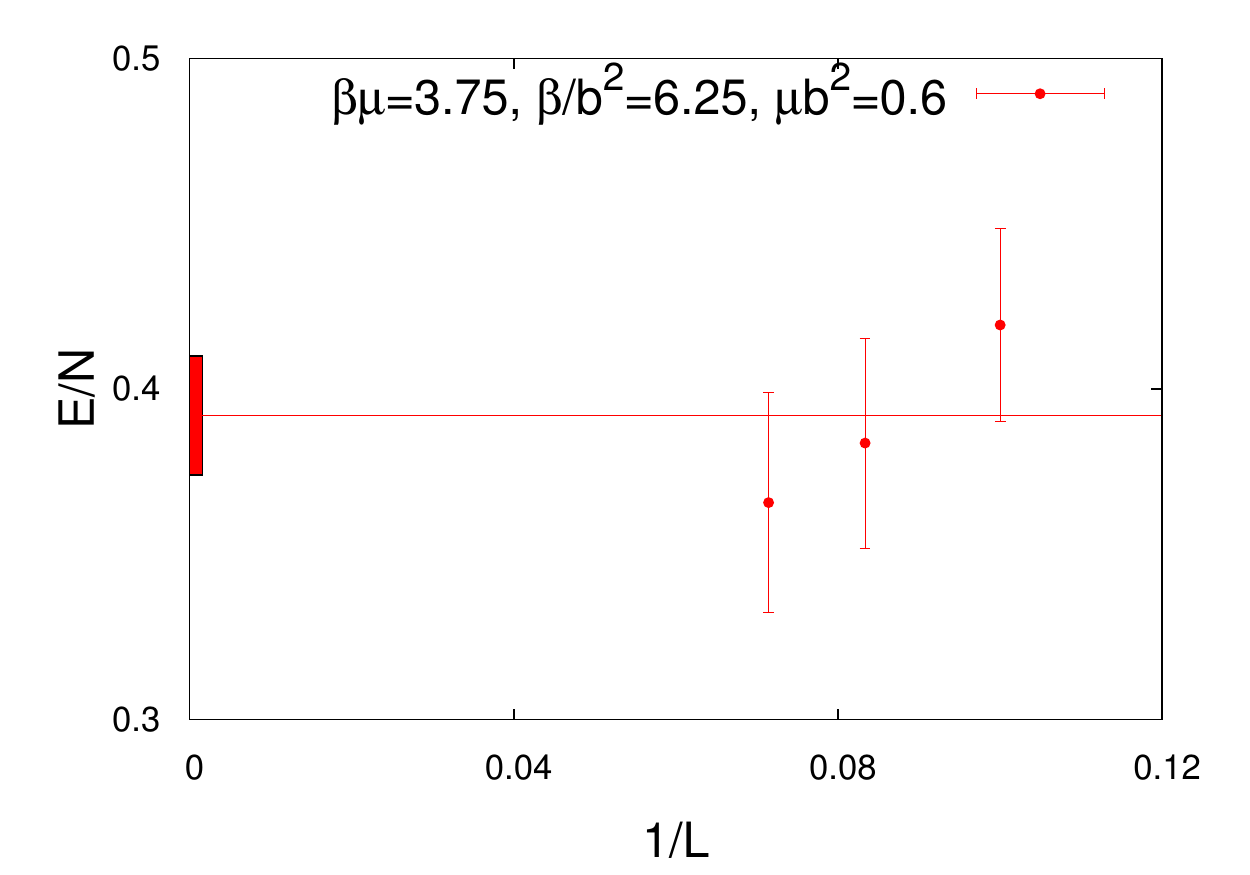}
\hspace{-4mm}\includegraphics[width=0.35\textwidth]{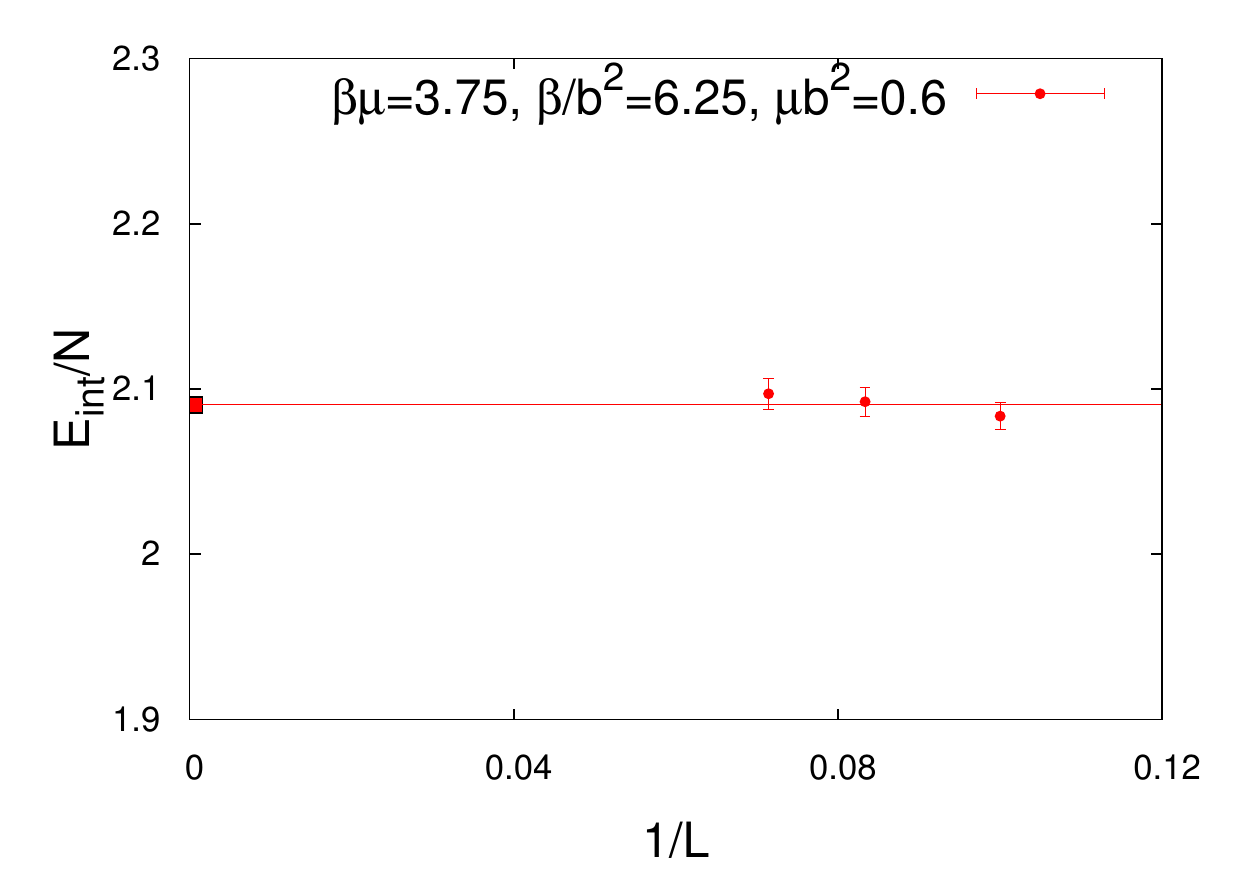}

\hspace{-5mm}\includegraphics[width=0.35\textwidth]{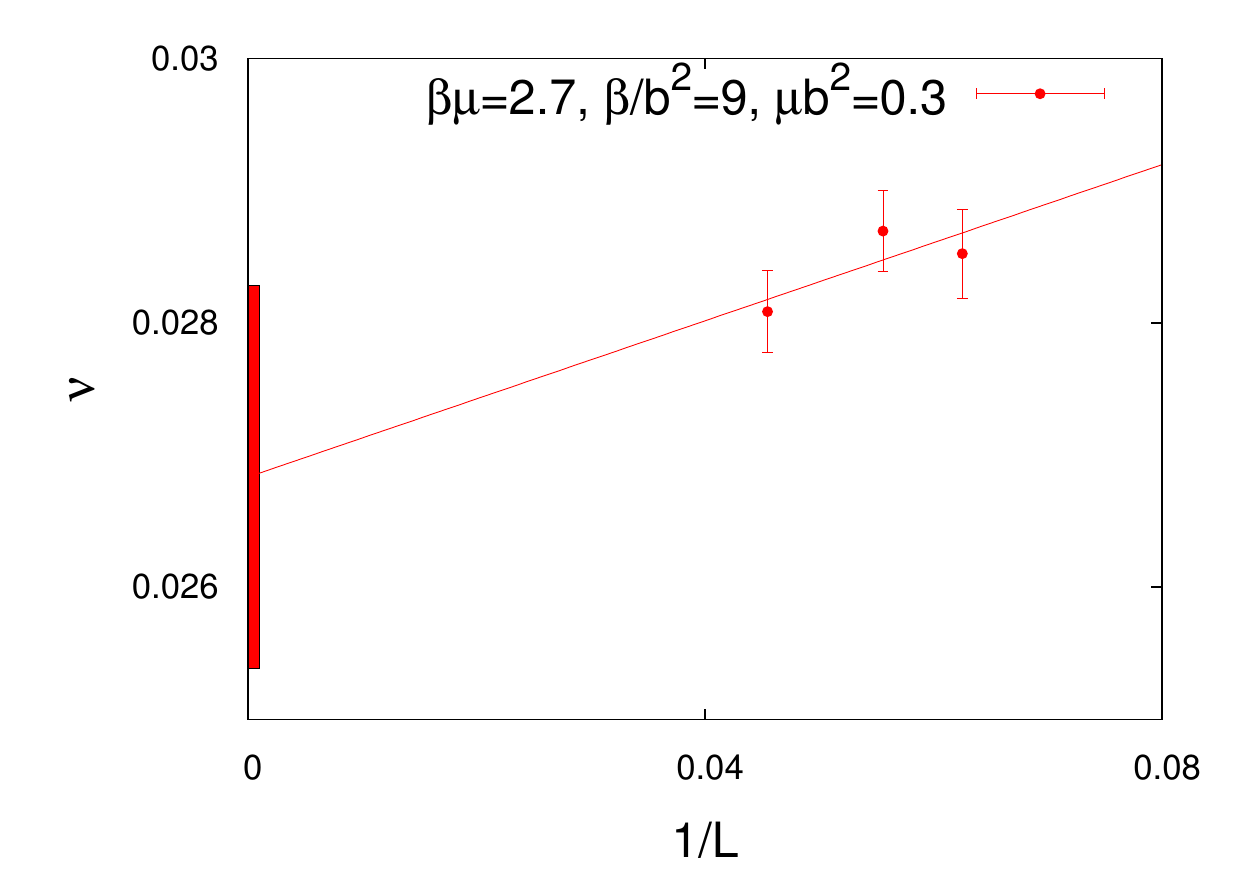}
\hspace{-4mm}\includegraphics[width=0.35\textwidth]{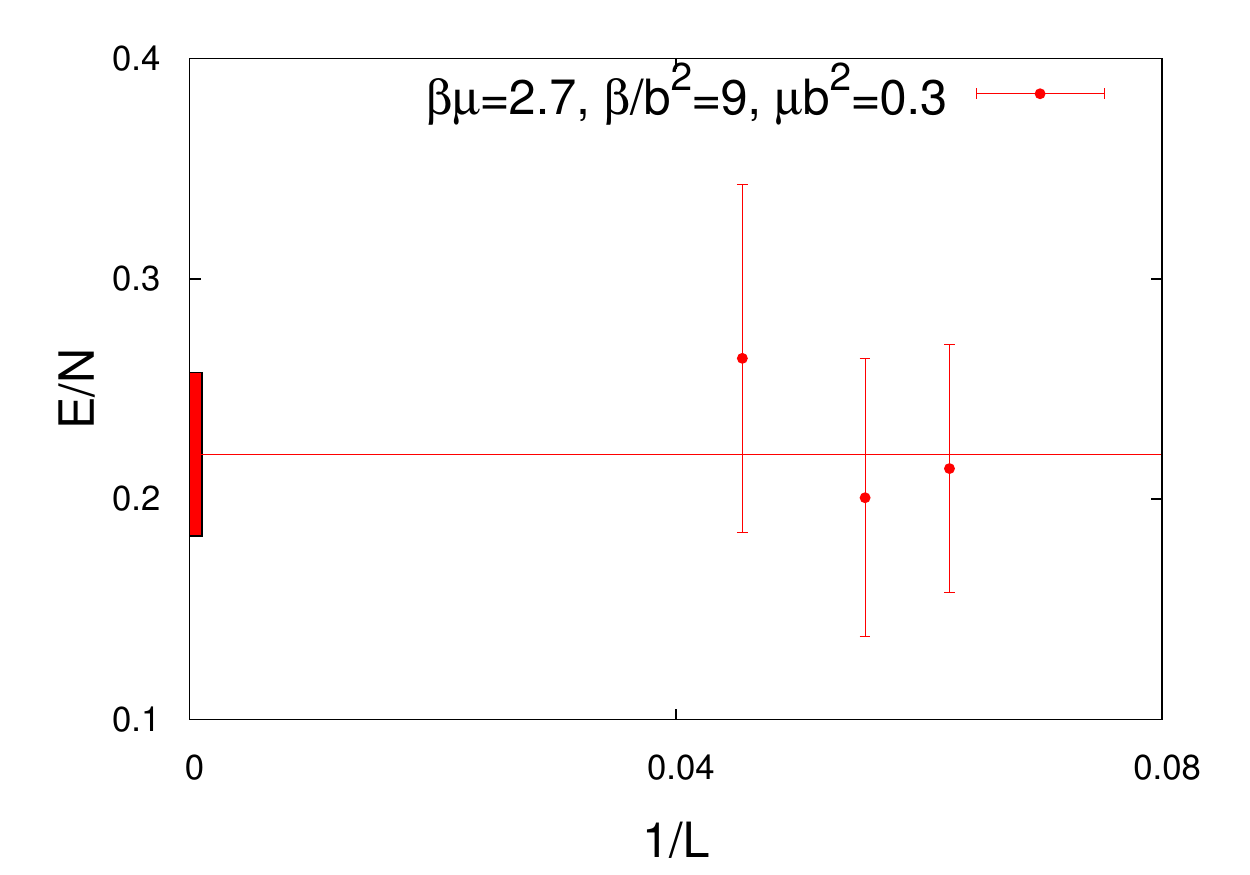}
\hspace{-4mm}\includegraphics[width=0.35\textwidth]{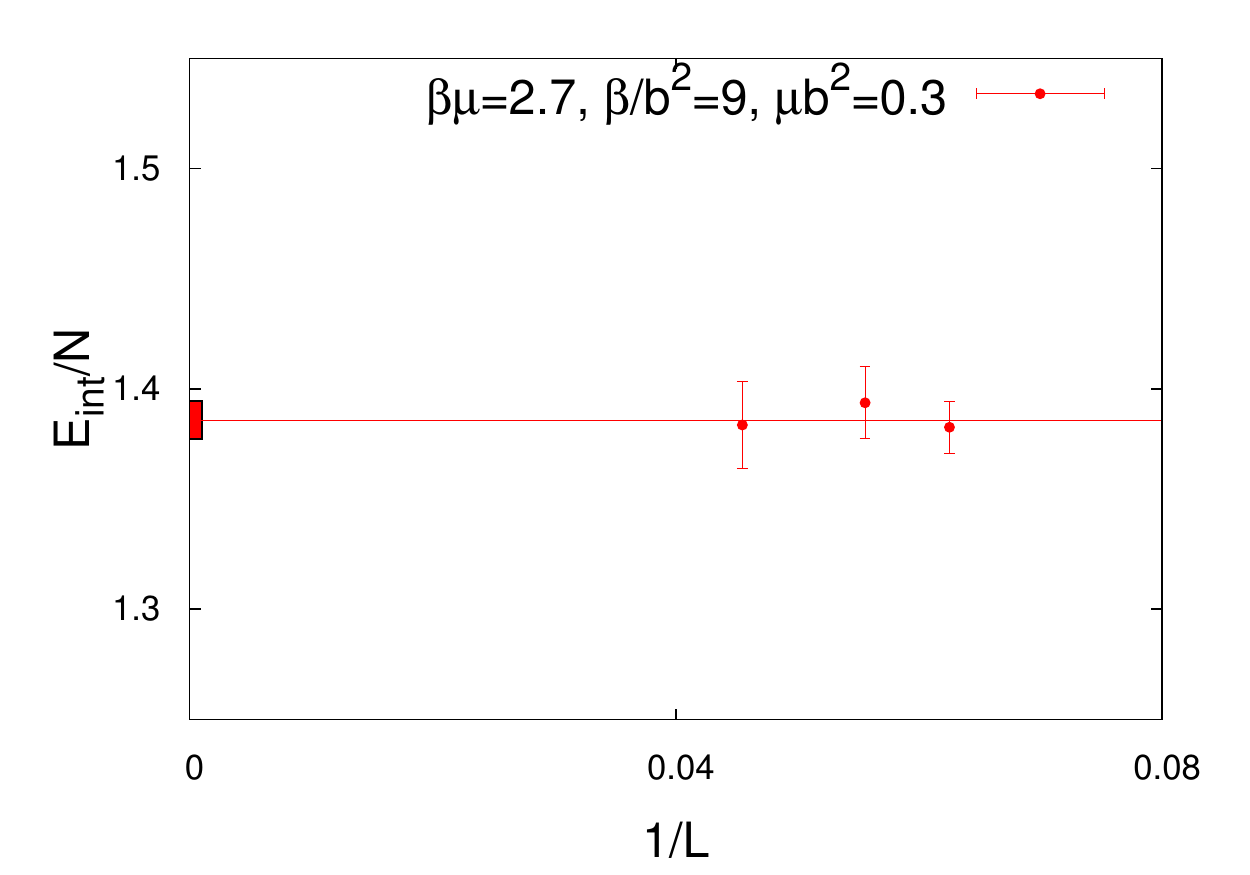}
\caption{\label{fig:TDextrapols_lowT}Examples of thermodynamic limit
  extrapolations in the superfluid phase for the filling factor $\nu$
  (left), the energy per particle $E/N=E/L^3\nu$ (middle), and the interaction
  energy per particle $E_{\rm int}/N=E_{\rm int}/L^3\nu$ (right), which yields
  the value of the contact.}
\end{figure*}

\begin{figure*}
\centering
\hspace{-5mm}\includegraphics[width=0.35\textwidth]{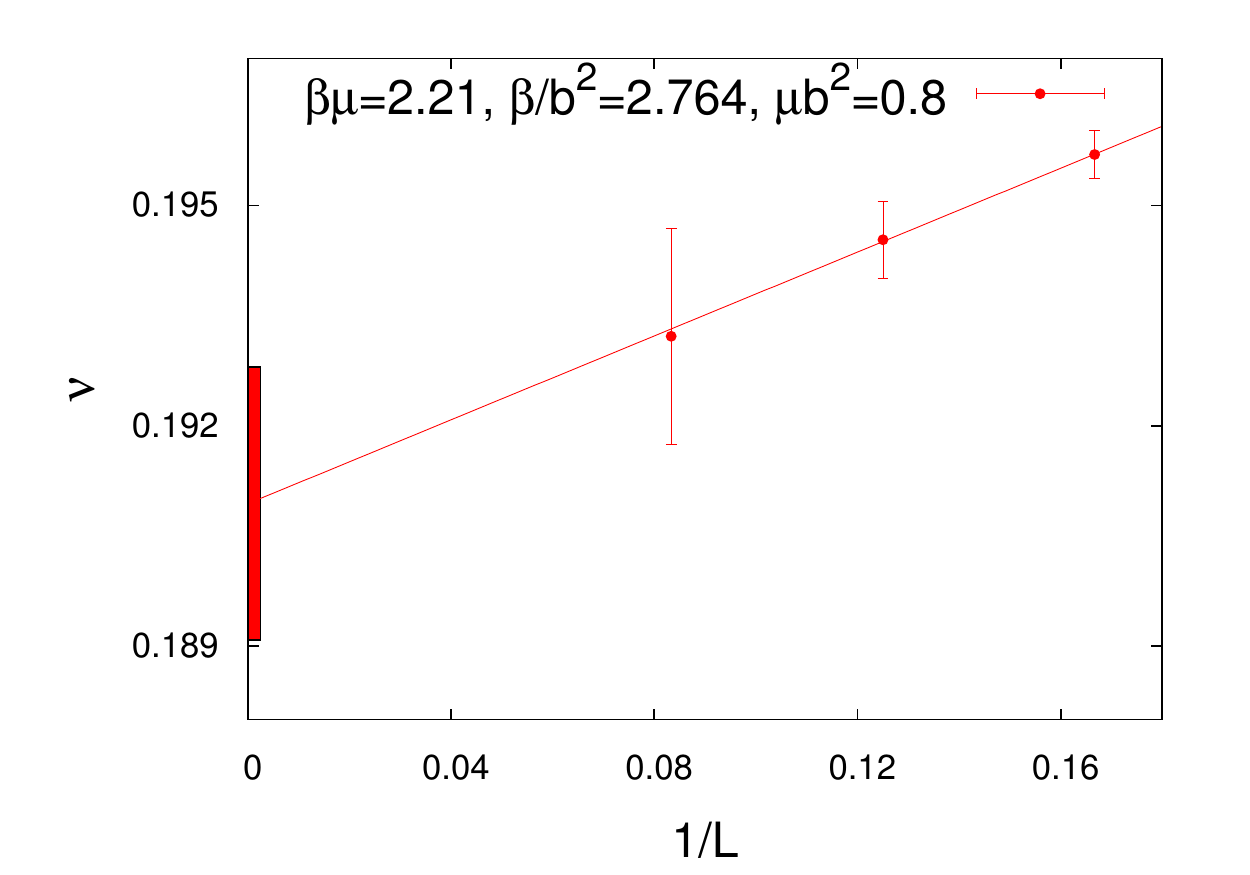}
\hspace{-4mm}\includegraphics[width=0.35\textwidth]{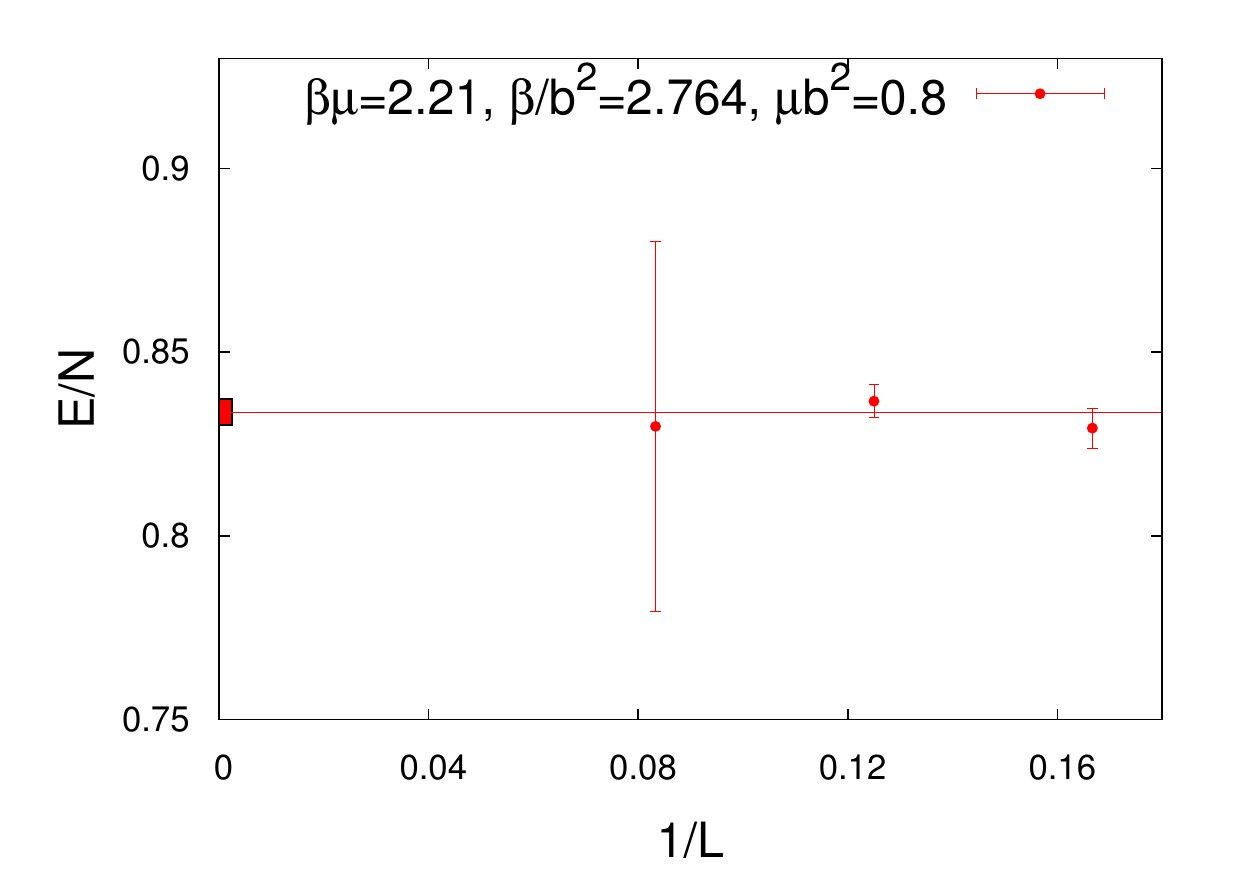}
\hspace{-4mm}\includegraphics[width=0.35\textwidth]{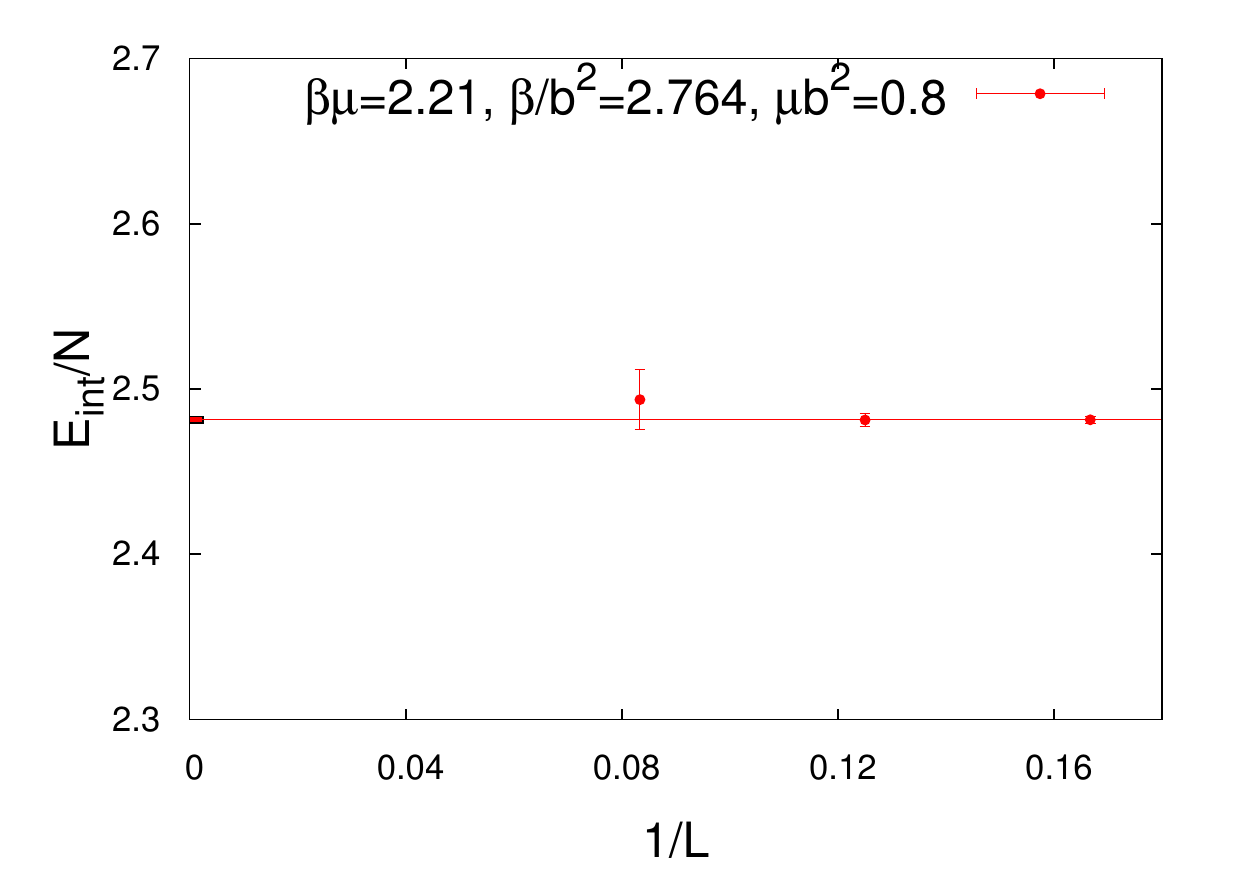}

\hspace{-5mm}\includegraphics[width=0.35\textwidth]{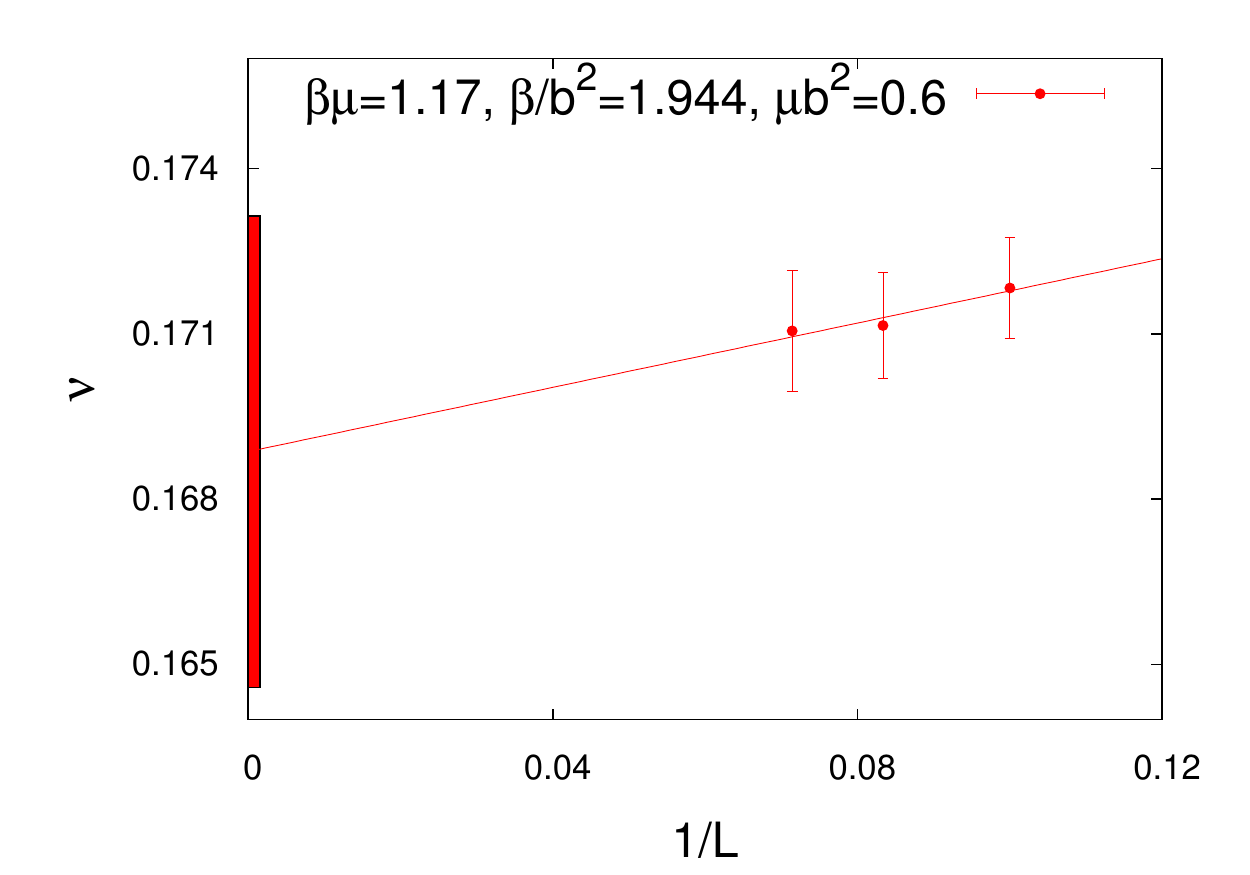}
\hspace{-4mm}\includegraphics[width=0.35\textwidth]{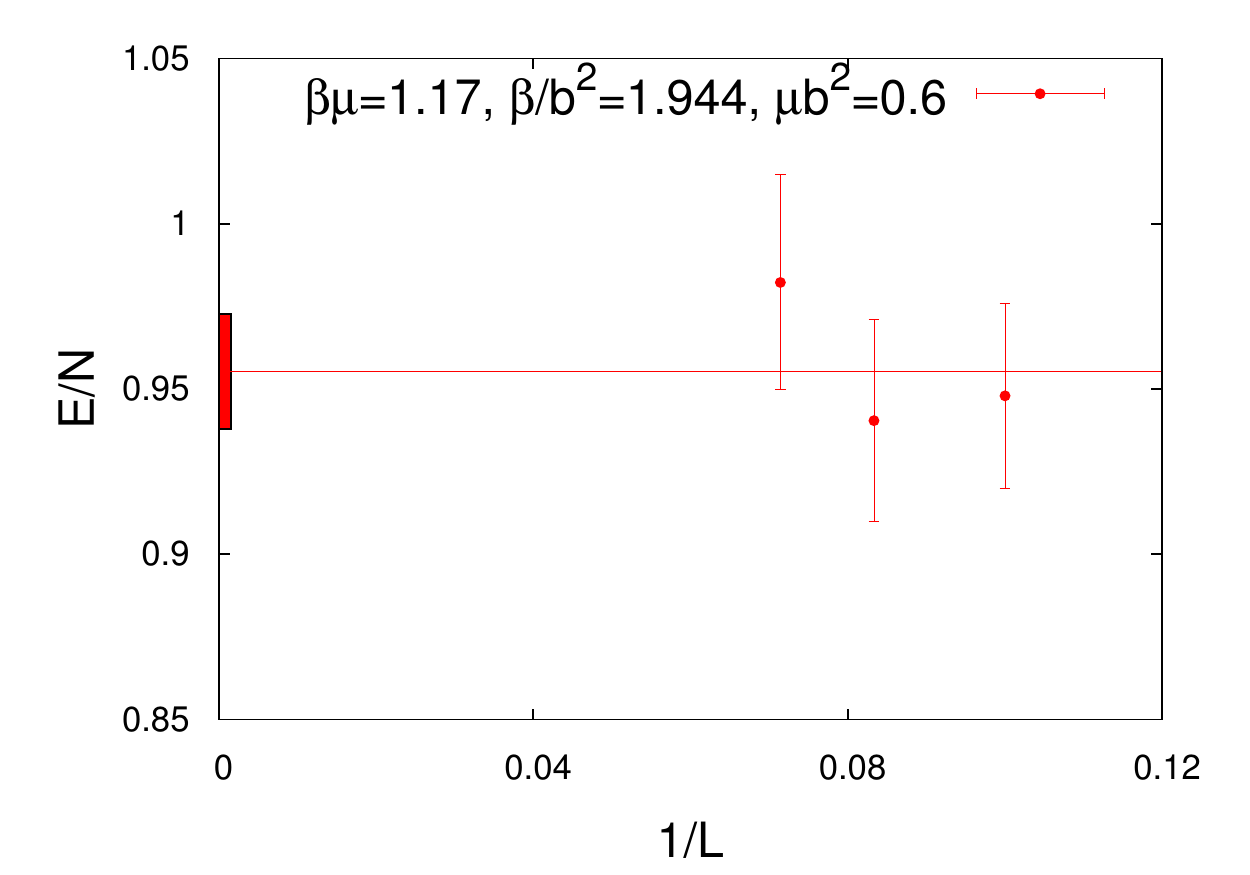}
\hspace{-4mm}\includegraphics[width=0.35\textwidth]{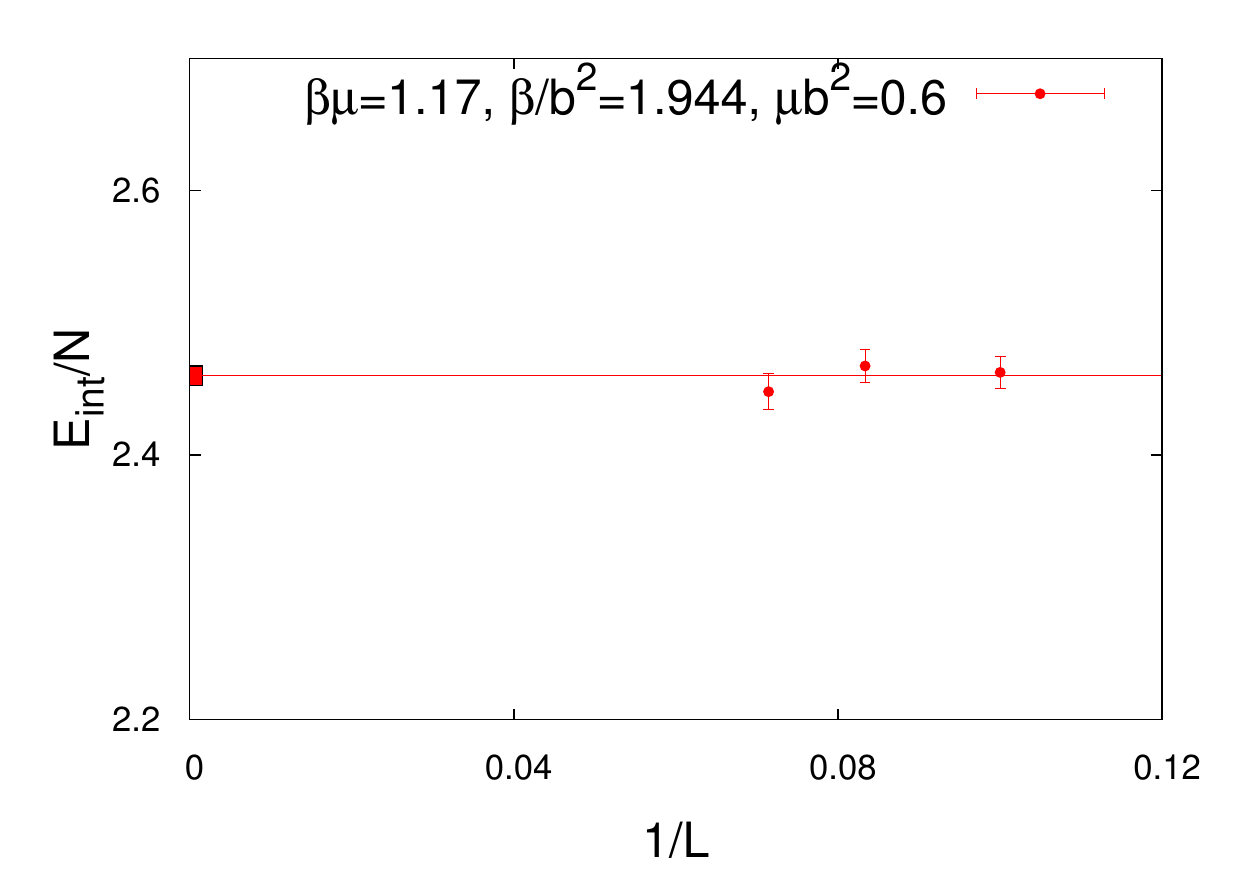}

\hspace{-5mm}\includegraphics[width=0.35\textwidth]{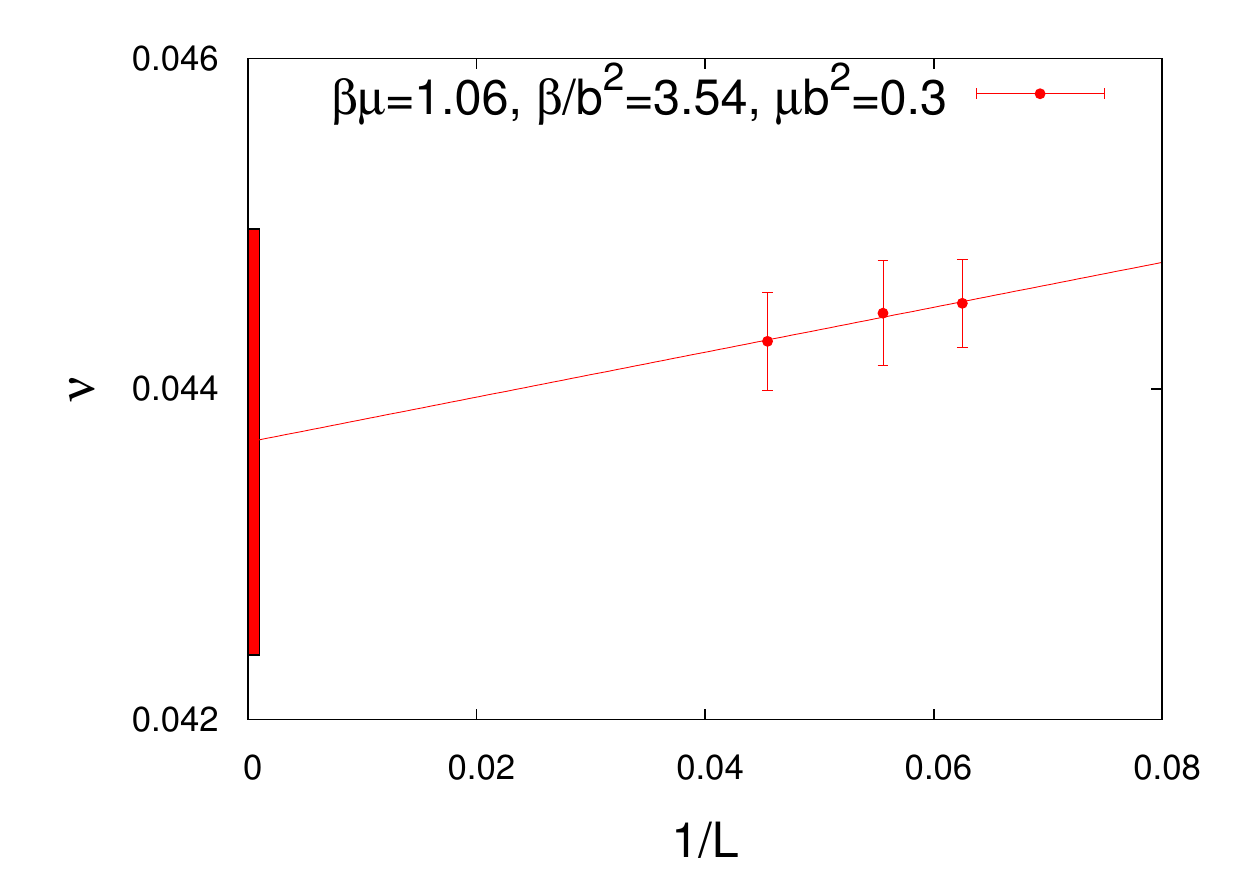}
\hspace{-4mm}\includegraphics[width=0.35\textwidth]{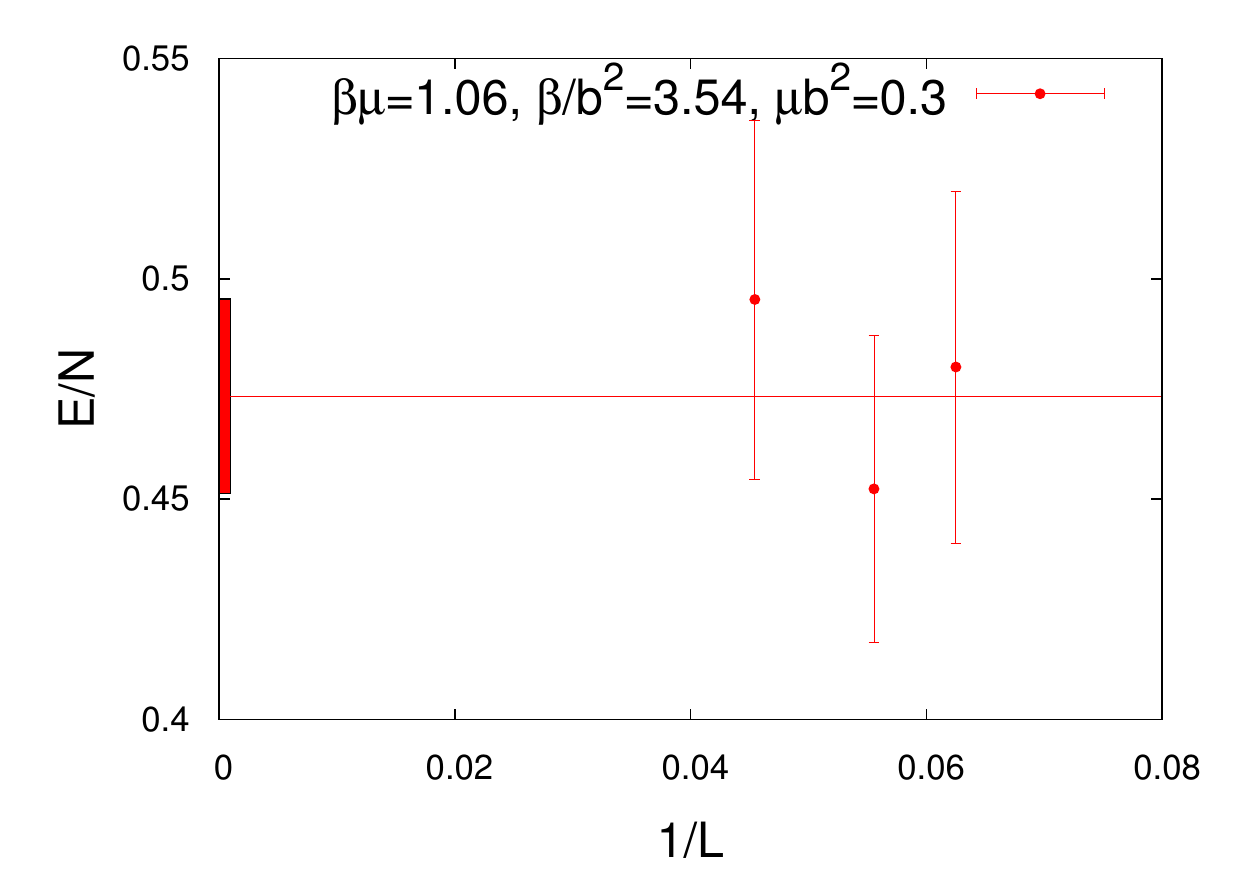}
\hspace{-4mm}\includegraphics[width=0.35\textwidth]{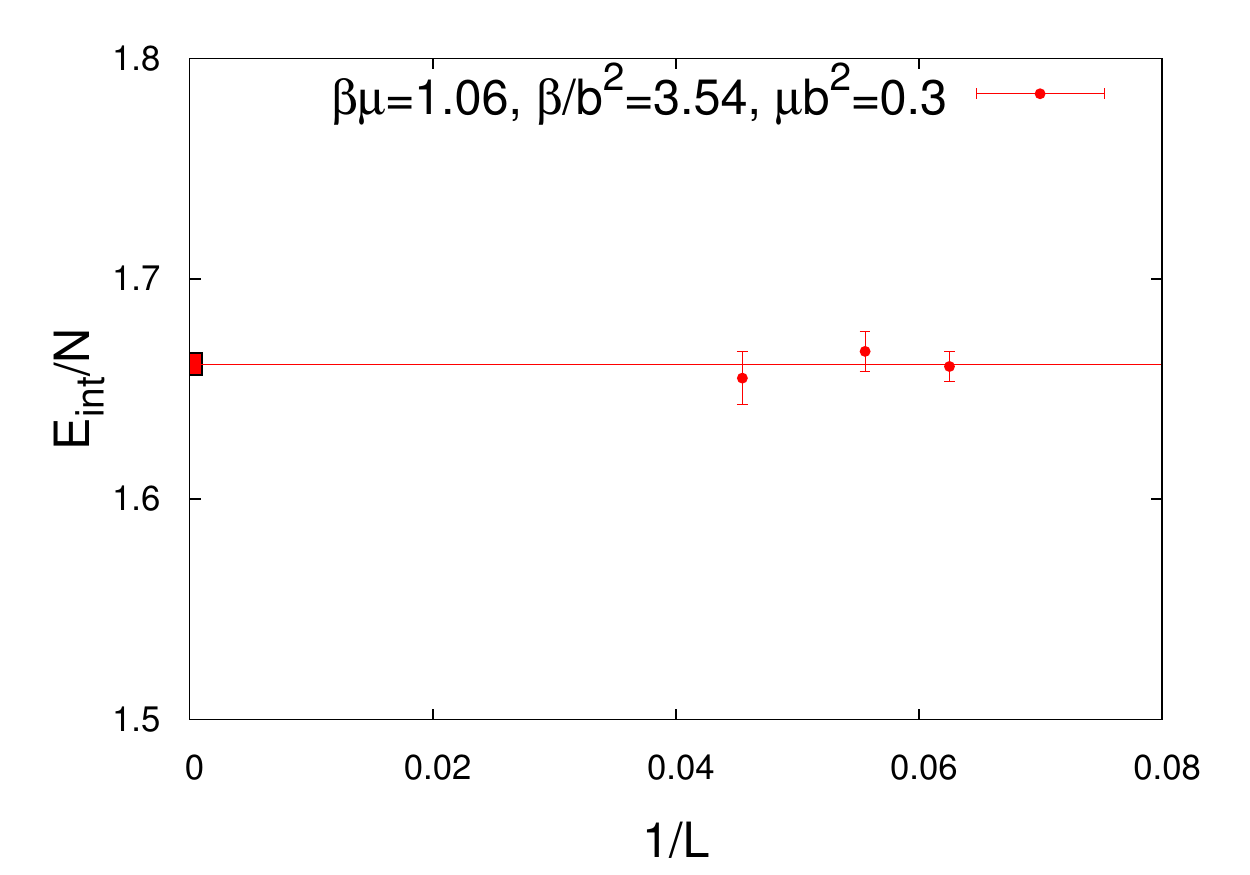}
\caption{\label{fig:TDextrapols_highT}Examples of thermodynamic limit extrapolations in the normal phase for the filling factor $\nu$ (left), the energy per particle $E/N=E/L^3\nu$ (middle), and the interaction energy per particle $E_{\rm int}/N=E_{\rm int}/L^3\nu$ (right), which yields the value of the contact.}
\end{figure*}

We set the physical scale via $\nu=nb^3$, where $\nu=\langle\sum_{\sigma}c^\dagger_{\mathbf{x}\sigma}c_{\mathbf{x}\sigma}\rangle$ is the dimensionless filling factor and $n$ the particle number density. Due to universality all physical quantities are given in units which can be expressed as appropriate powers of the Fermi energy $\ef=(3\pi^2 n)^{2/3}$. To extract the physical results we need to perform two limits: the thermodynamic limit to infinite system size and then the continuum limit to zero lattice spacing.

First we take the thermodynamic limit in order to estimate and reduce
systematic errors due to finite volume. For each $(\mu b^2,\beta/b^{2})$
we perform computations with several (usually 3 or 4) sizes of cubic volumes,
$V = L^3$, and extrapolate results as $1/L \to 0$. At the lowest filling
factor (smallest $b$) we used volumes up to $V=32^3$ (corresponding to up to
1000 particles). Typical ranges are $L=6$ to $L=14$ at higher filling factors
(about 40 to 700 particles) and $L=10$ to $L=32$ at lower filling factors
(about 100 to 1000 particles). We find that the filling factor $\nu$ is the
quantity most sensitive to finite-volume effects. With periodic boundary
  conditions, we expect finite-volume effects to cause an increase in filling
  factor compared to the infinite-volume limit.  In small volumes the
  particles will feel an enhanced attraction not just from their neighboring
  particles, but also from their round-the-world doppelg\"angers.
In agreement with \cite{burovski} we observe that the data for the filling
factor are fit well by a linear function of $1/L$. Once we have $\nu$ in the thermodynamic limit, we can obtain $\ef$ and the
 dimensionless observables by taking $\nu$ to the appropriate power and
 multiplying by quantities which have no statistically significant finite-volume errors. Within the statistical uncertainties, it appears that finite-volume errors are negligible for $E/N$ and $E_{\mathrm{int}}/N$.
Examples of thermodynamic extrapolations are shown in 
Figs.~\ref{fig:TDextrapols_lowT} and \ref{fig:TDextrapols_highT}
for different parameter sets in the superfluid and normal phases, respectively.

For the continuum limit we vary the dimensionless chemical potential $\mu b^2$
such that the filling factor tends to zero. This is equivalent to
$b\rightarrow0$ since $b\propto\nu^{1/3}$ if $n$ is fixed to be a constant,
physical value. Dimensionless ratios of physical quantities can then
be extrapolated to the continuum limit by assuming discretization effects
can be parametrized using a power series in $b$, or equivalently $\nu^{1/3}$.

Our previous work \cite{ourmain, ourproc2} focused on determining the critical
temperature and computing thermodynamic observables there. The critical
temperature was determined at several values of the lattice spacing. In
practice this was done by varying the inverse temperature and chemical
potential in lattice units to the point where a finite-size scaling study
of the order parameter indicated the transition would occur in infinite
volume.  Here we extend that work by taking the continuum limit of observables
for a range of temperatures on either side of the phase transition.

Ideally the continuum limit would be taken varying $b$ along lines of constant $\beta\mu$. The numerical data acquired,
however, do not lie exactly along lines of constant $\beta\mu$.
Therefore, we group the data in several narrow bands of $\betamu =
\beta \mu$ values and extrapolate the data within each band to the
continuum limit. Each band is defined by a central value $\betamu_0$
and a width, as shown in Fig.~\ref{fig:betamu_nu13}. In order to
account for mild $y$ dependence, we find it sufficient to introduce a
term proportional to $\delta \betamu = \betamu - \betamu_0$ in the
extrapolation function.  The lattice spacing dependence of a physical
quantity $X$ can be written as a power series in the lattice spacing
$b \propto \nu^{1/3}$ \cite{Symanzik:1983dc}. Therefore, our continuum
limit fits are to functions of the form
\begin{align}
X(\betamu_0; \nu, \delta\betamu) &= X_0 \left( 1 + d_1 \delta\betamu 
+ \sum_{k=1}^K c_k \nu^{k/3} \right) \,.
\label{eq:continuum}
\end{align} 
The fit parameter $X_0 = X_0(\betamu_0)$ is then taken to be the continuum limit
result. The other fit parameters, $d_1$ and $c_k$ also depend on $\betamu_0$, but
we suppress this dependence in the notation.  In almost every case the Monte
Carlo data are sufficient to determine $c_1$ but not the coefficients of
higher-order terms. In other words, the data points indeed look linear in $\nu^{1/3}$. However,
given that, especially for larger $\betamu$, the numerical values of $\nu^{1/3}$ 
are not very small, it is prudent to allow for higher-order contributions
in the numerical data.  Therefore we introduce Bayesian prior distributions
for the $c_k$ with $k>1$ \cite{Lepage:2001ym}.  In the cases where the Monte Carlo results do constrain
$c_2$ (i.e.\ fits to $\mu/\ef$ at low $\betamu$) we find $c_2
\approx 0.3$.  Therefore, we take Gaussian prior distributions centered
at 0 with width 0.3.  We found very little difference in the fits where we
set $K=2$ or $K=3$, but we used the latter, more conservative, option for the
results presented here.
Finally, we also performed fits which included a term $f_1 \,\delta\betamu \,
\nu^{1/3}$, with a Gaussian prior for $f_1$ of $0.0 \pm 1.0$.  This had no
effect on the fits, so for simplicity we omit this term from our final fits.

\begin{figure*}
\centering
\includegraphics[width=0.32\textwidth]{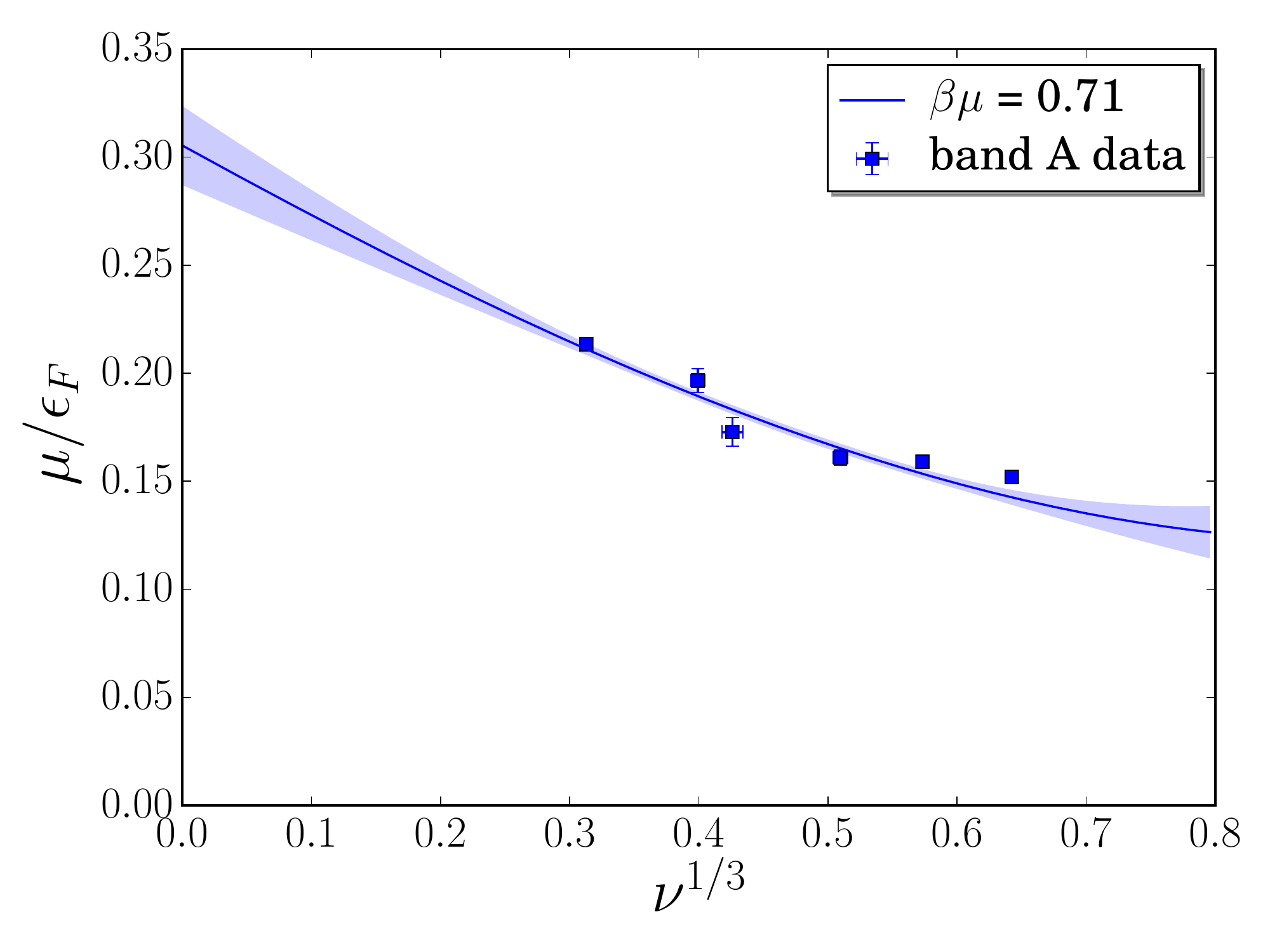}
\includegraphics[width=0.32\textwidth]{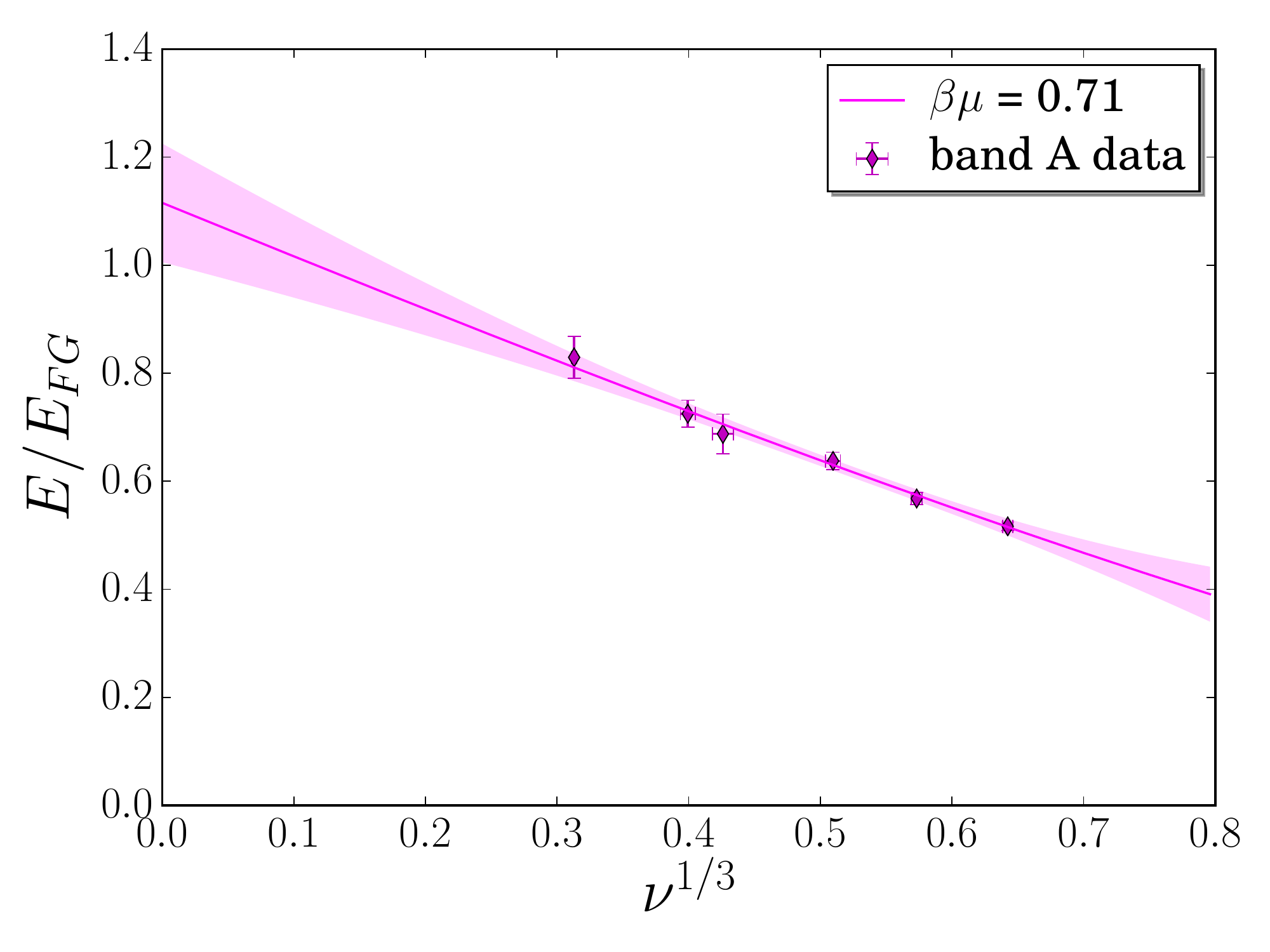}
\includegraphics[width=0.32\textwidth]{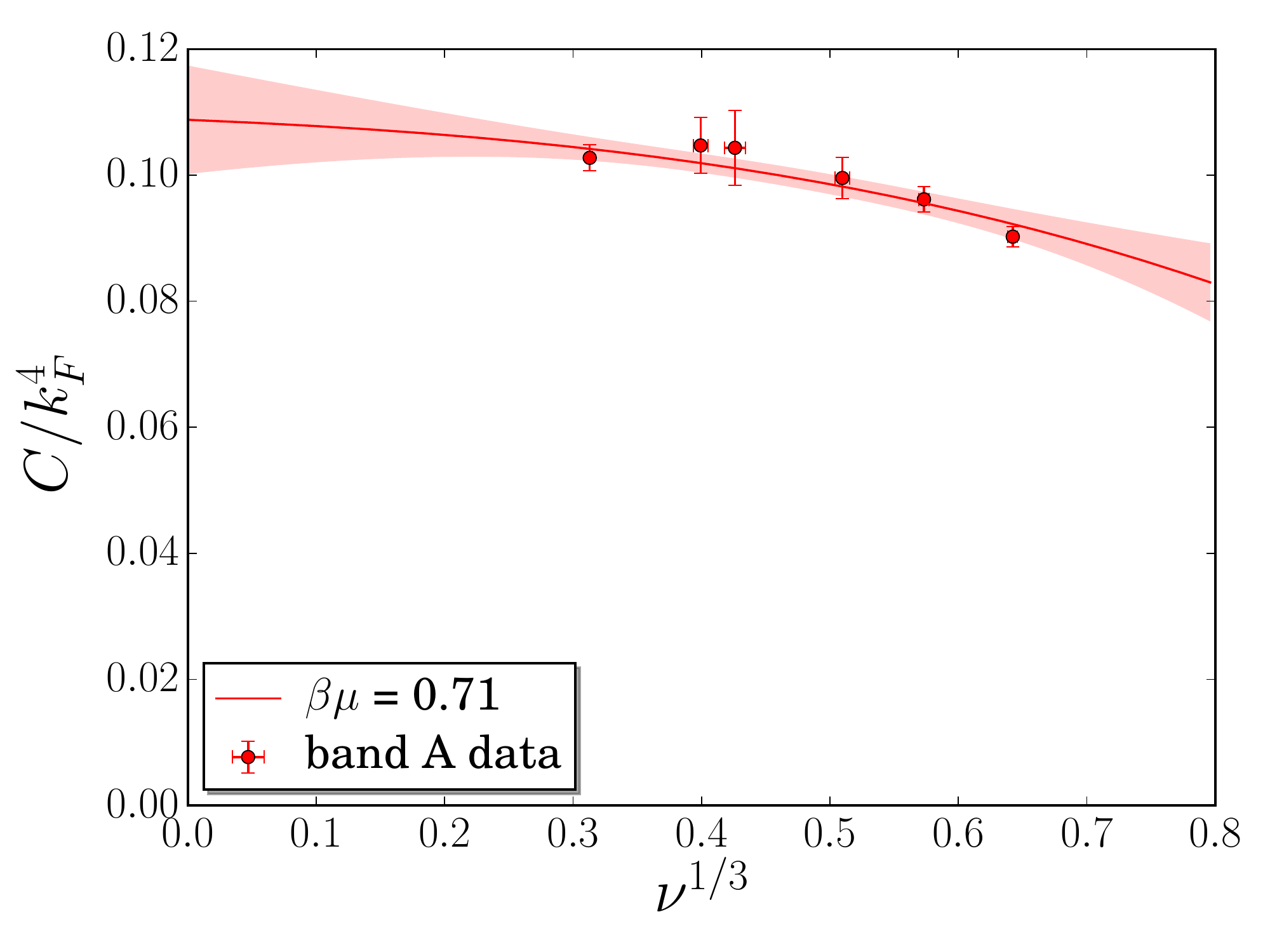}
\includegraphics[width=0.32\textwidth]{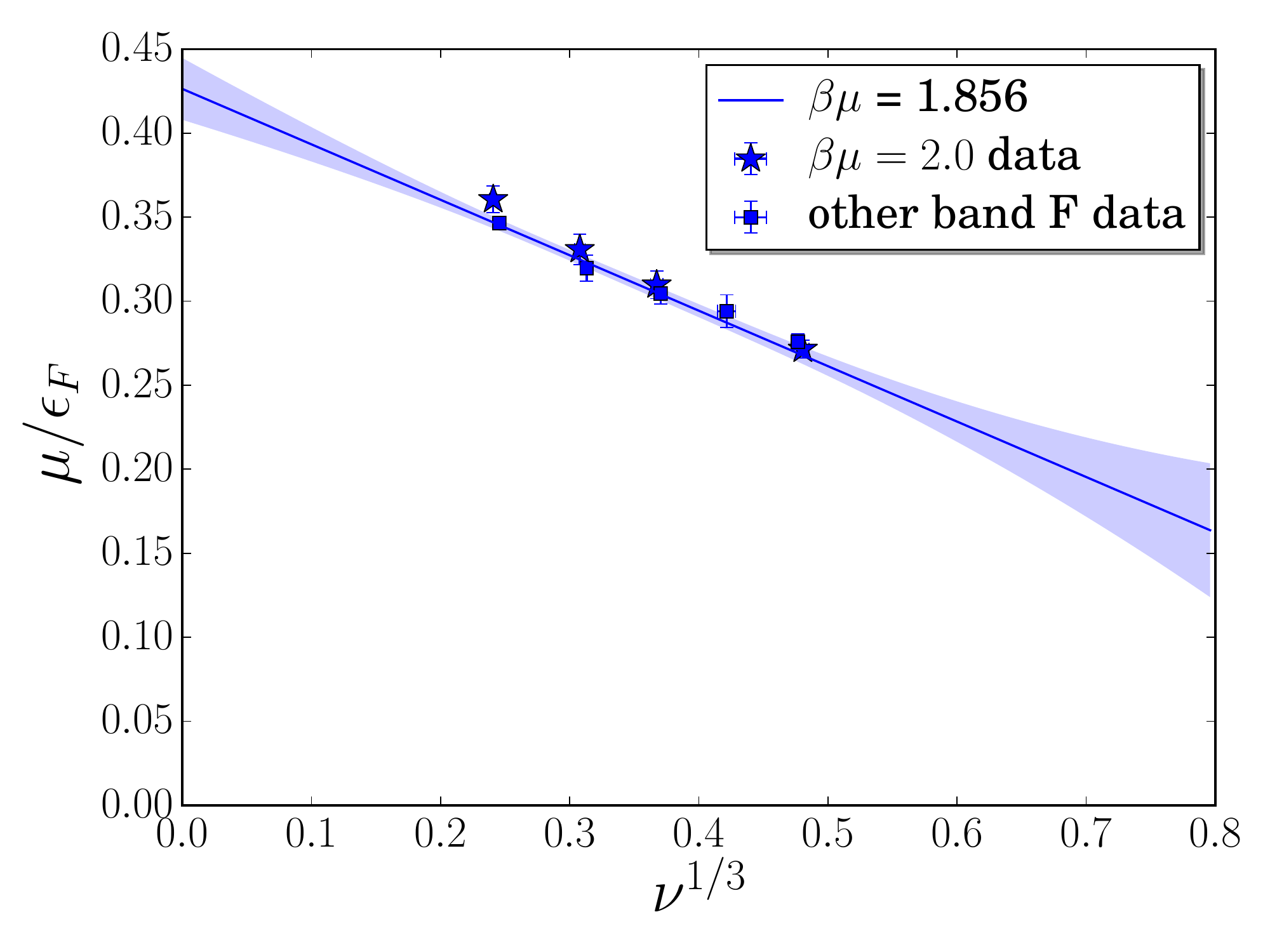}
\includegraphics[width=0.32\textwidth]{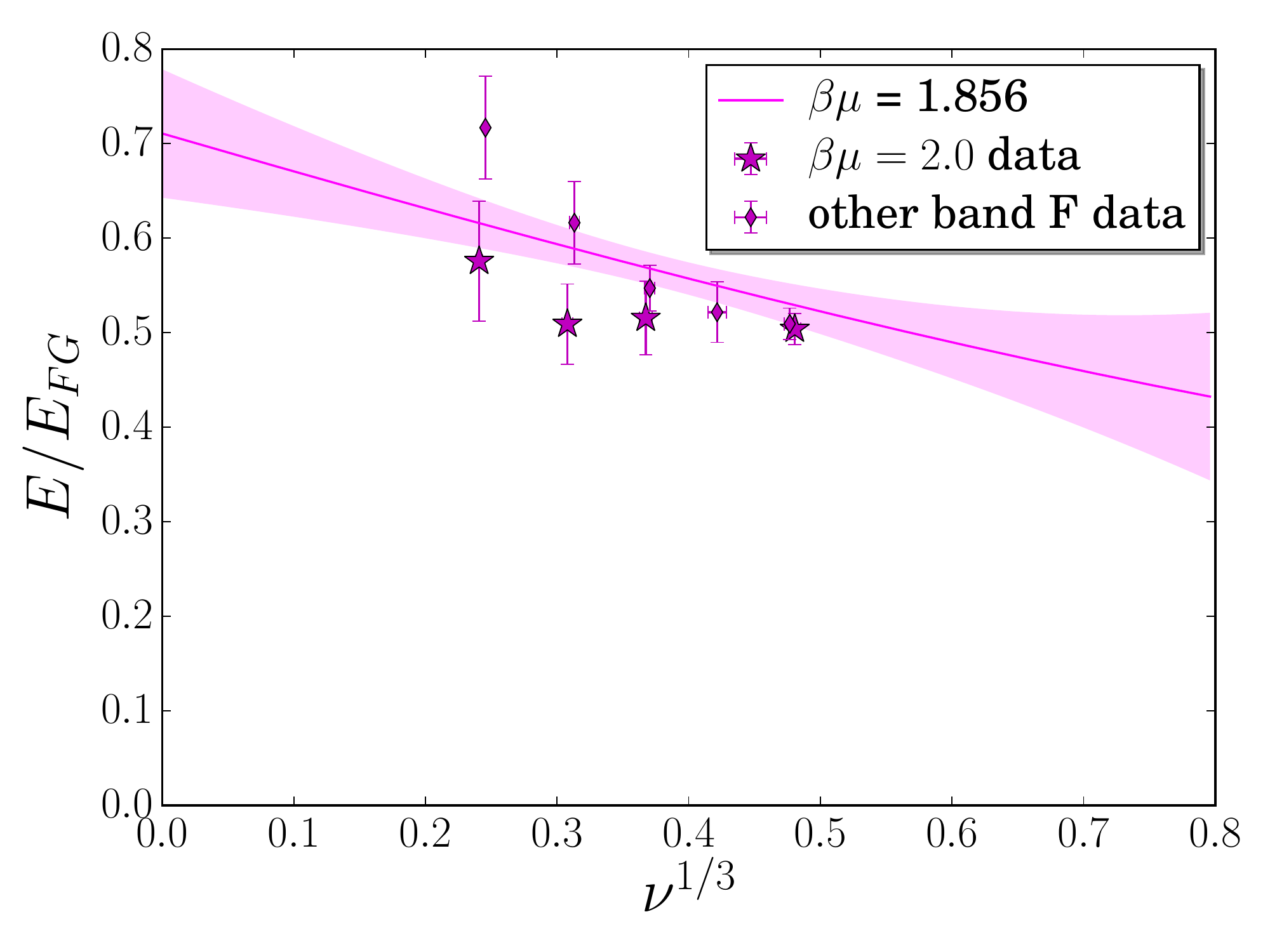}
\includegraphics[width=0.32\textwidth]{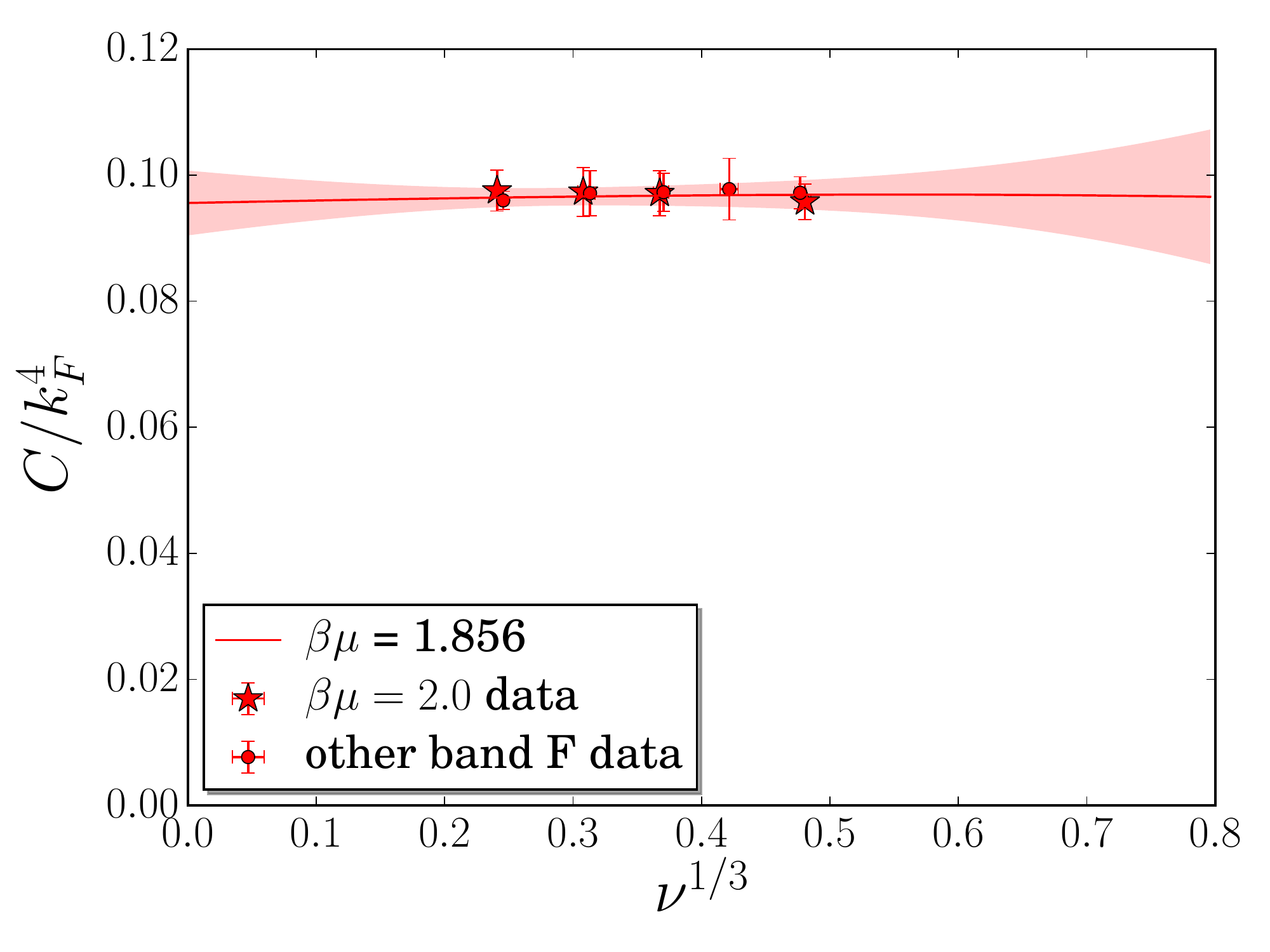}
\includegraphics[width=0.32\textwidth]{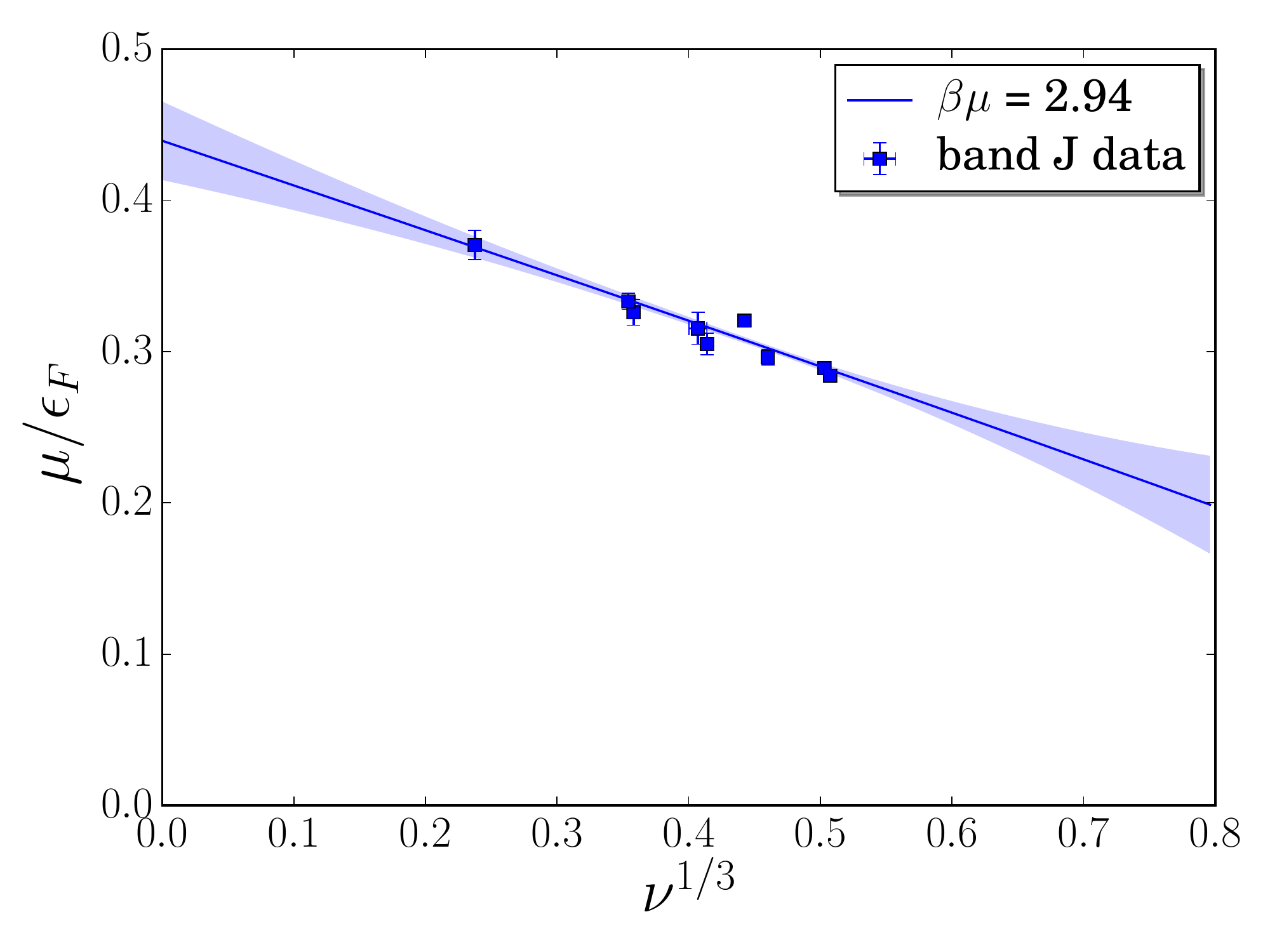}
\includegraphics[width=0.32\textwidth]{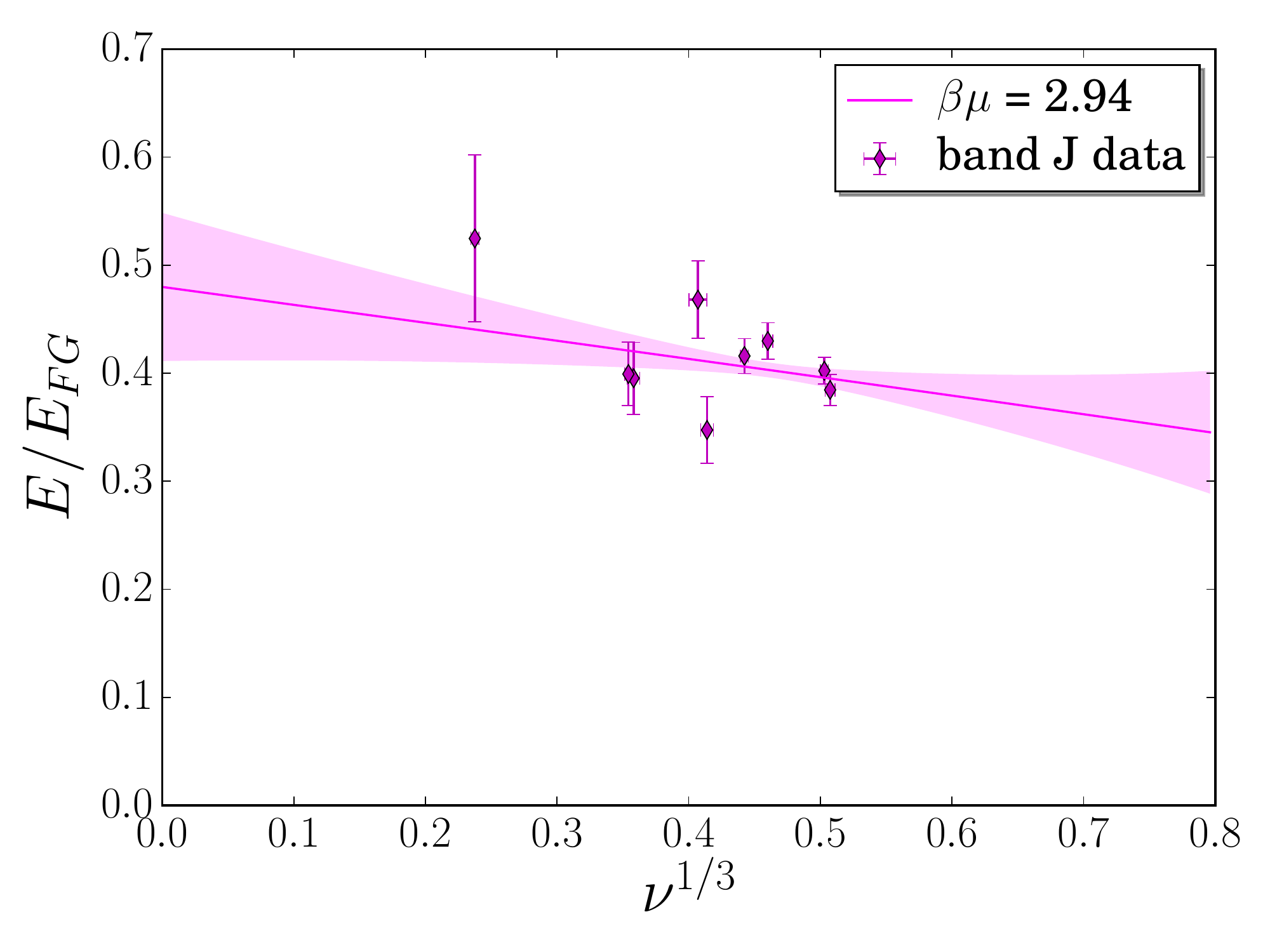}
\includegraphics[width=0.32\textwidth]{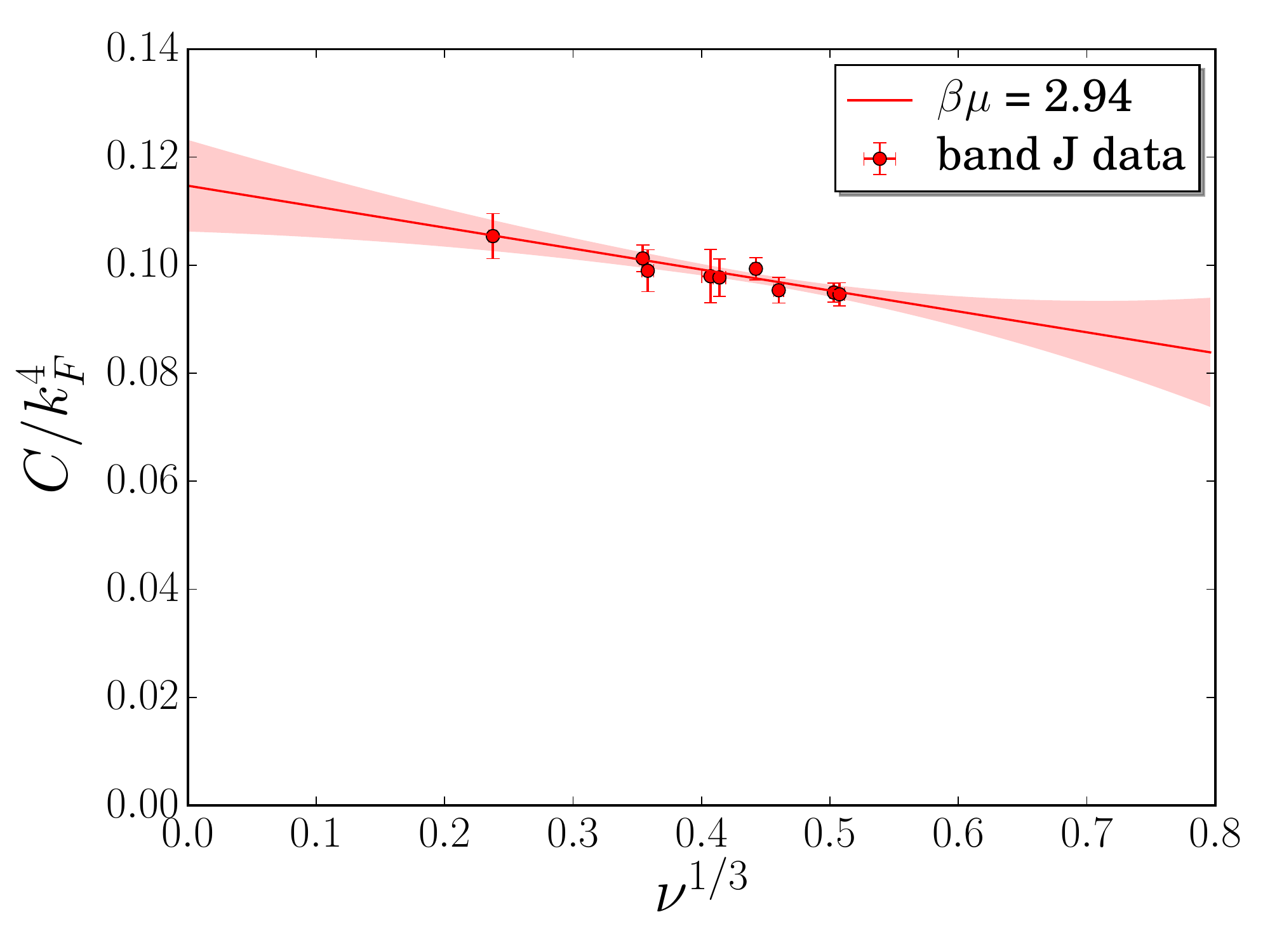}
\includegraphics[width=0.32\textwidth]{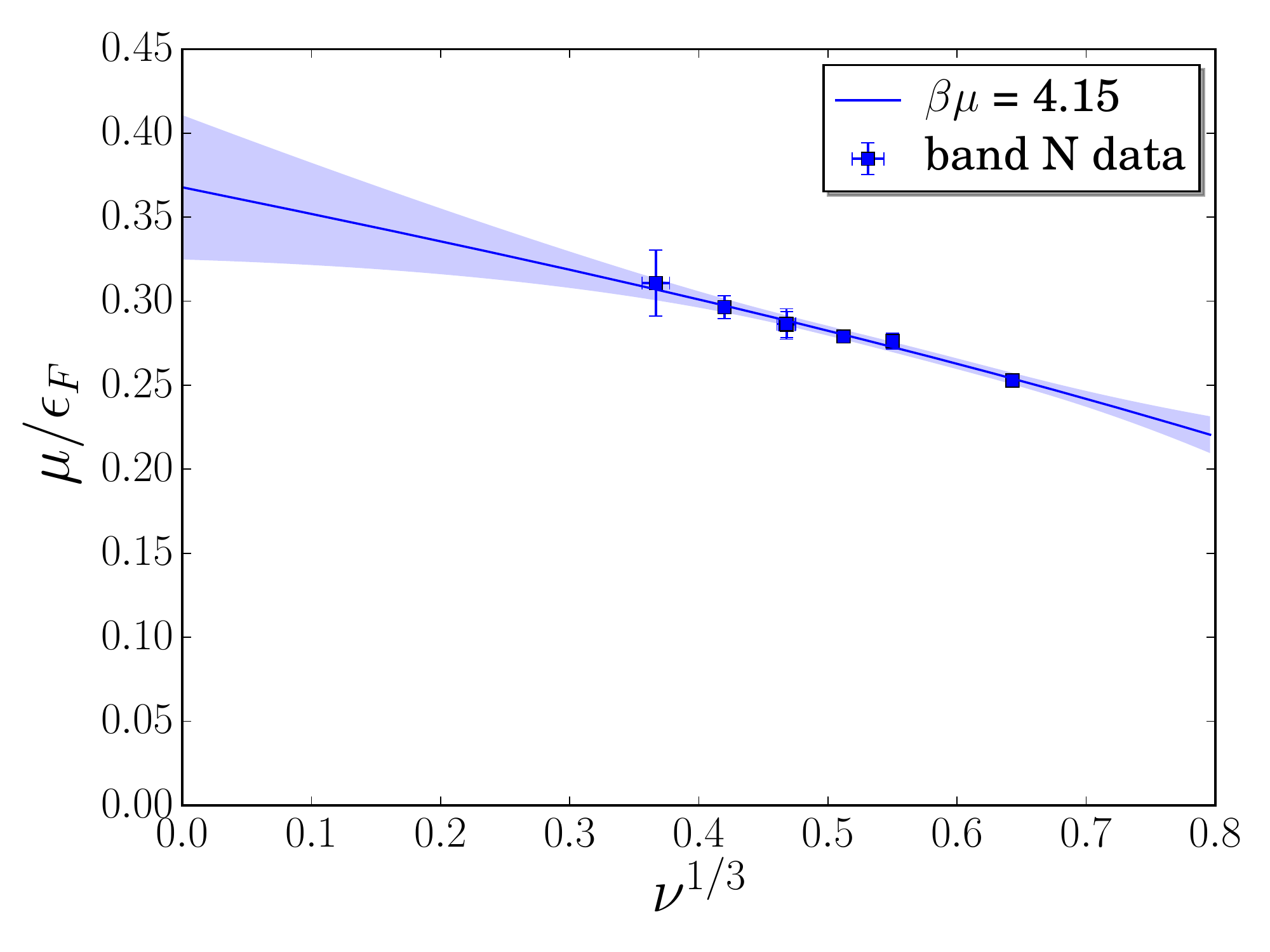}
\includegraphics[width=0.32\textwidth]{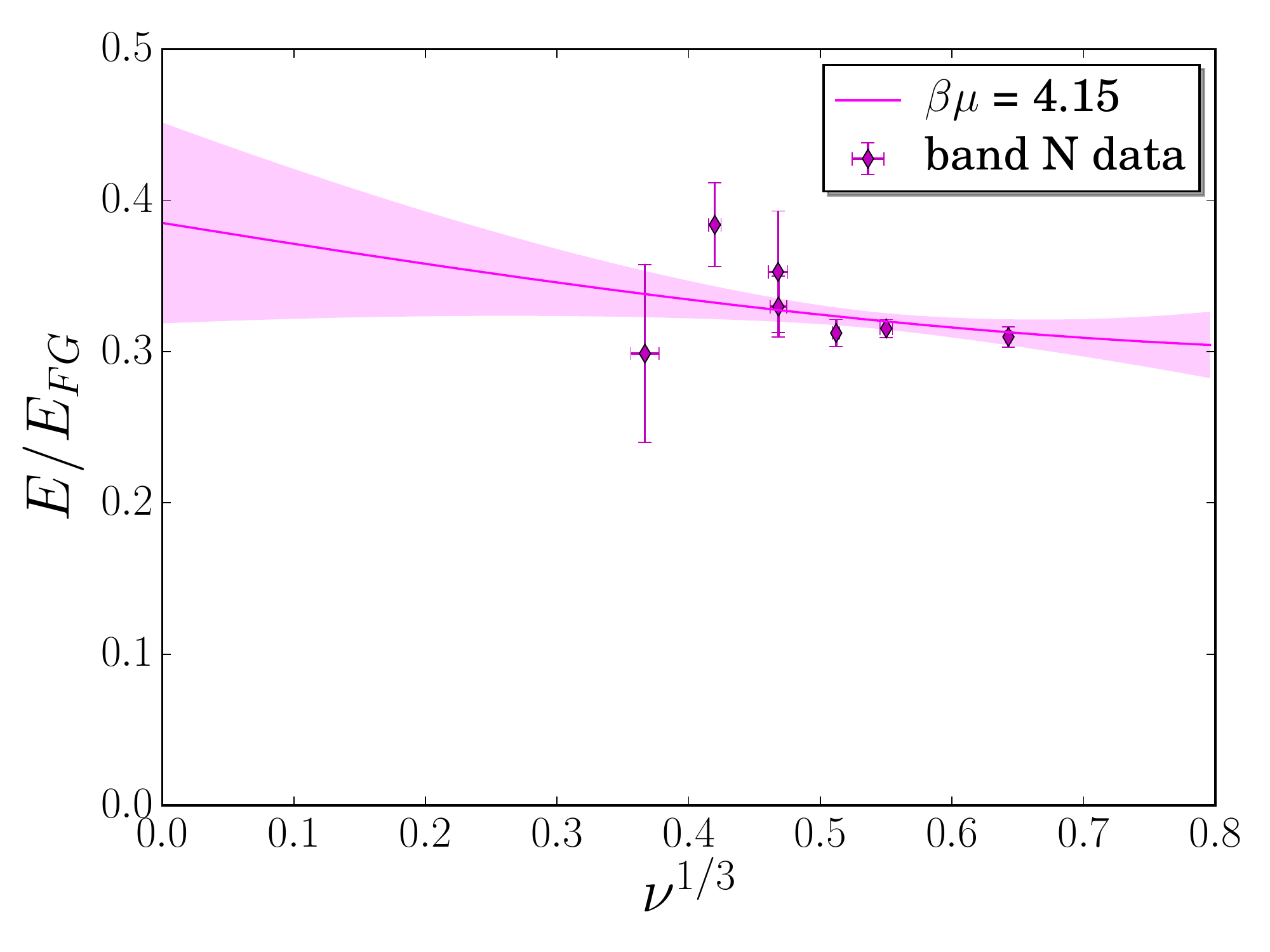}
\includegraphics[width=0.32\textwidth]{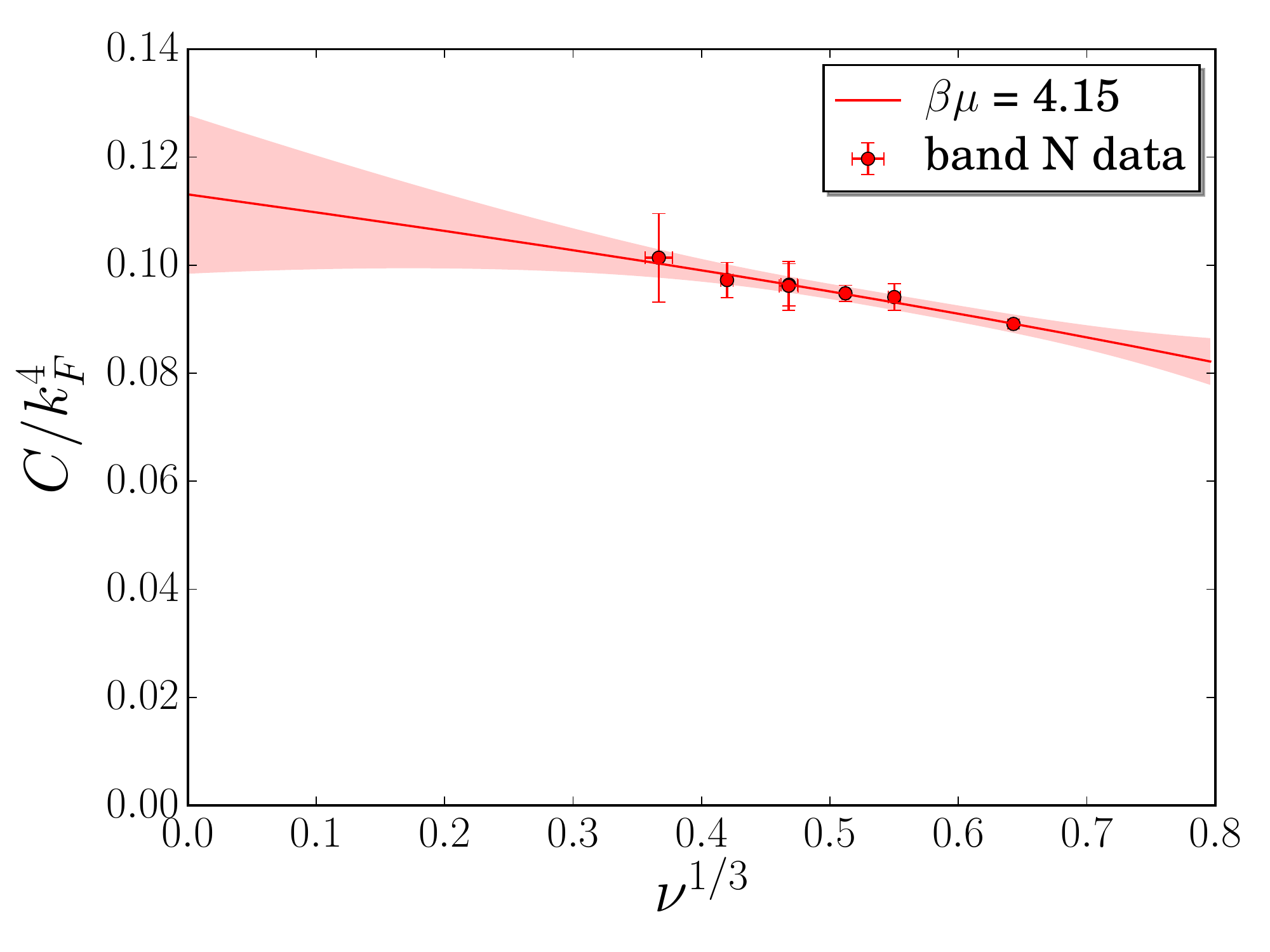}
\includegraphics[width=0.32\textwidth]{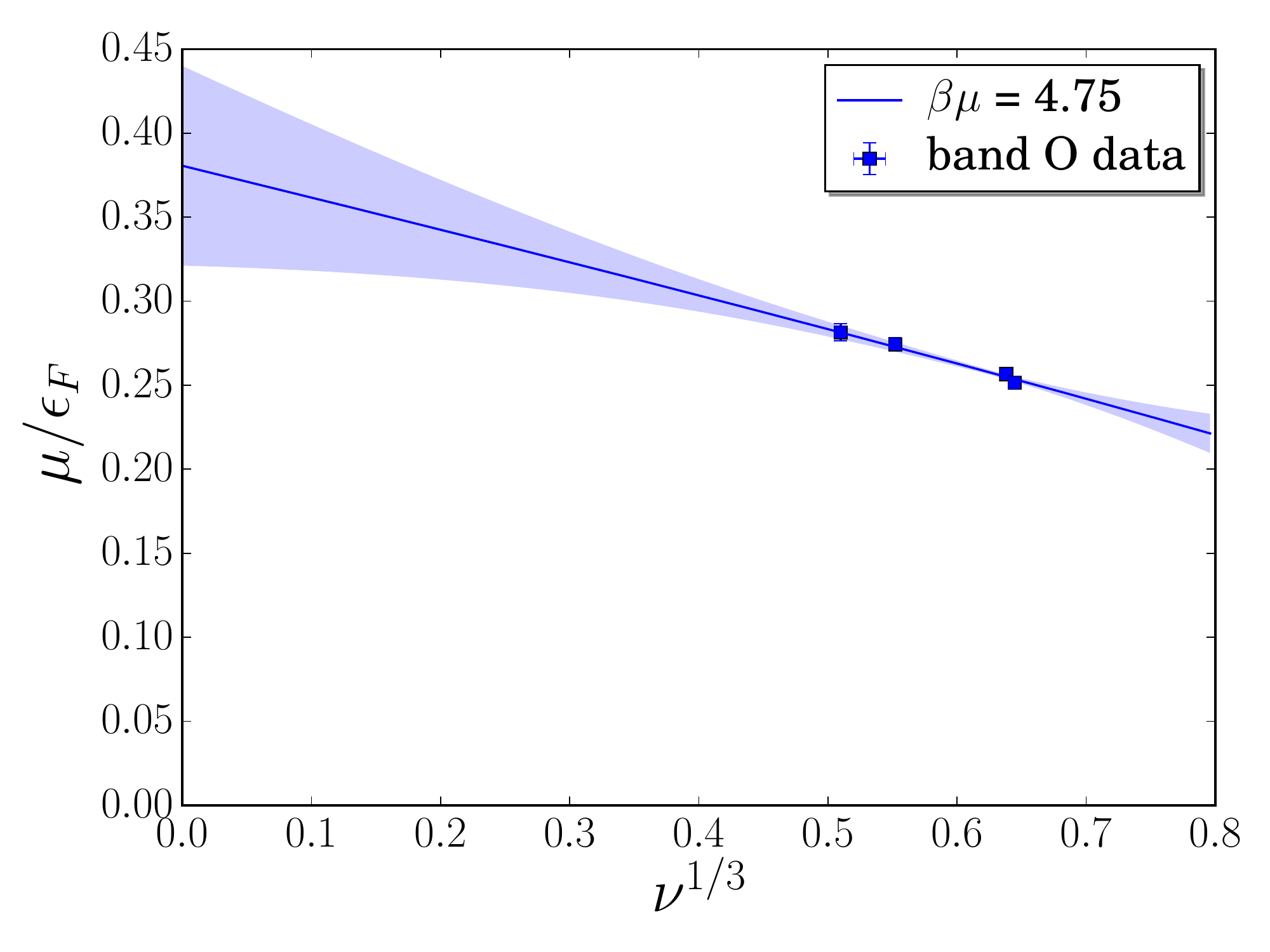}
\includegraphics[width=0.32\textwidth]{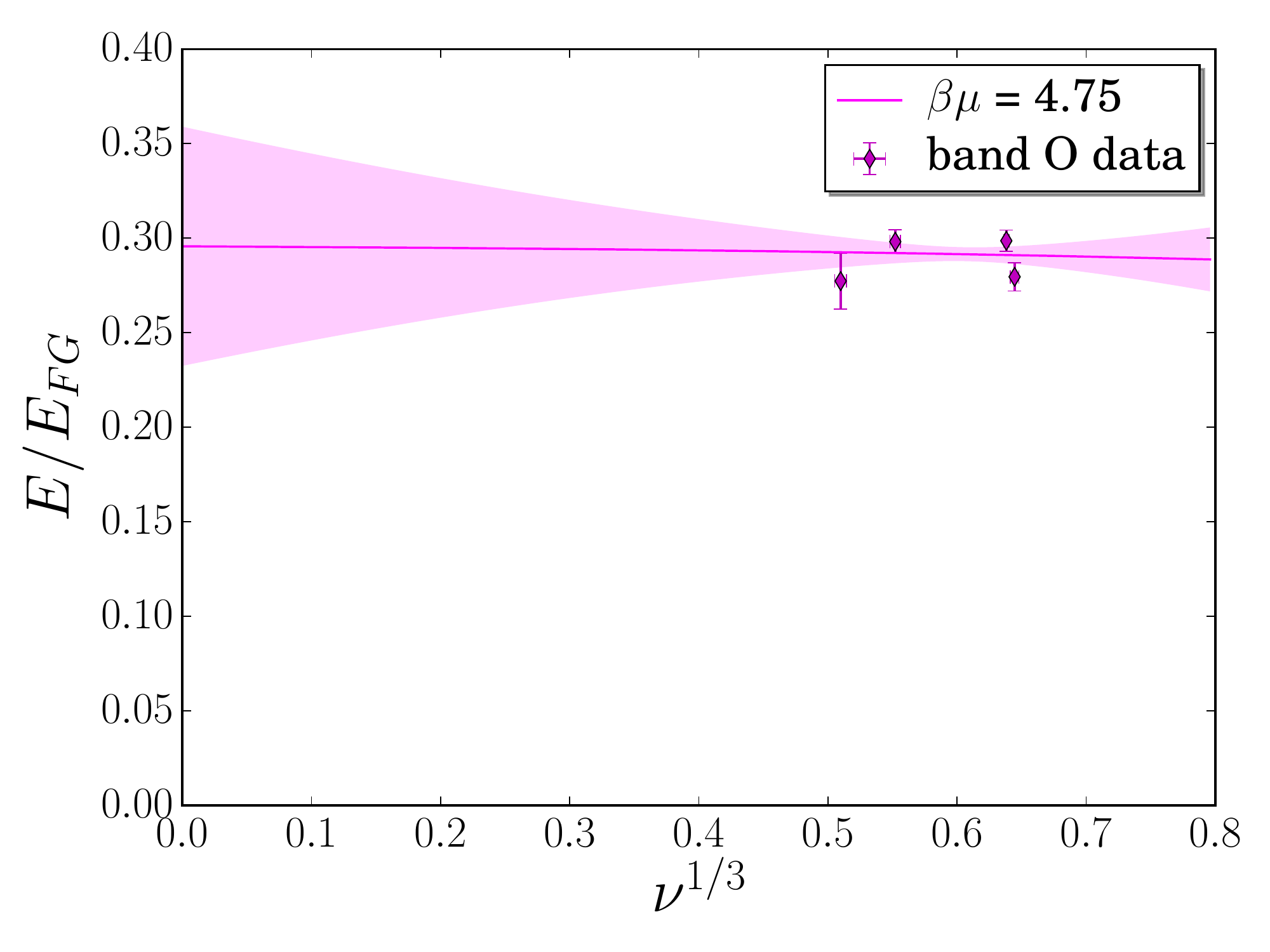}
\includegraphics[width=0.32\textwidth]{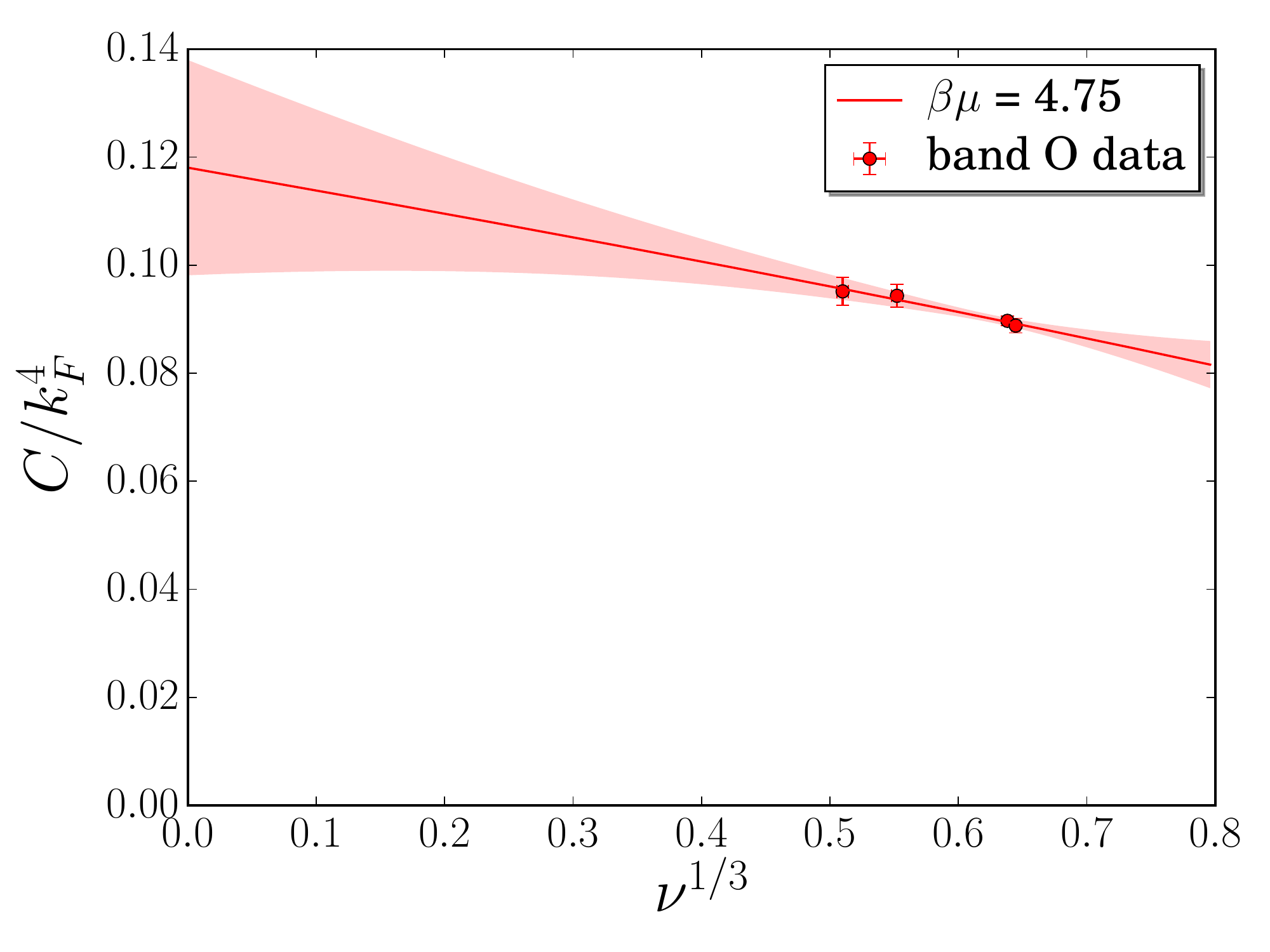}
\caption{\label{fig:continuum}Continuum limit extrapolation along bands 
 of constant $\beta\mu \pm \delta(\beta\mu)$ (see Fig.~\ref{fig:betamu_nu13}) 
for the chemical potential (left), energy 
density (middle) and contact density (right). See further discussion
at the end of Sec.~\ref{sec:limits} regarding the interpolation in $\beta\mu$.}
\end{figure*}

In Fig.~\ref{fig:continuum} we show the results of these fits for
5 of the 15 bands.  The fit curve and corresponding error are evaluated 
at the value of $\beta\mu$ given in the legend, and represent an interpolation
of the data to the central value of the lettered band shown in Fig.~\ref{fig:betamu_nu13} as well as the continuum extrapolation to $\nu =0$.  
In the case of band F, we have several data
points generated with $\beta\mu = 2$; we emphasize these points with stars.
Within uncertainties, the widths of the bands in $\beta\mu$ are sufficiently
narrow that interpolating in $\beta\mu$ is mild, well-parametrized by 
the $d_1$ term in Eq.~(\ref{eq:continuum}).

\begin{figure}
\centering
\includegraphics[width=\columnwidth]{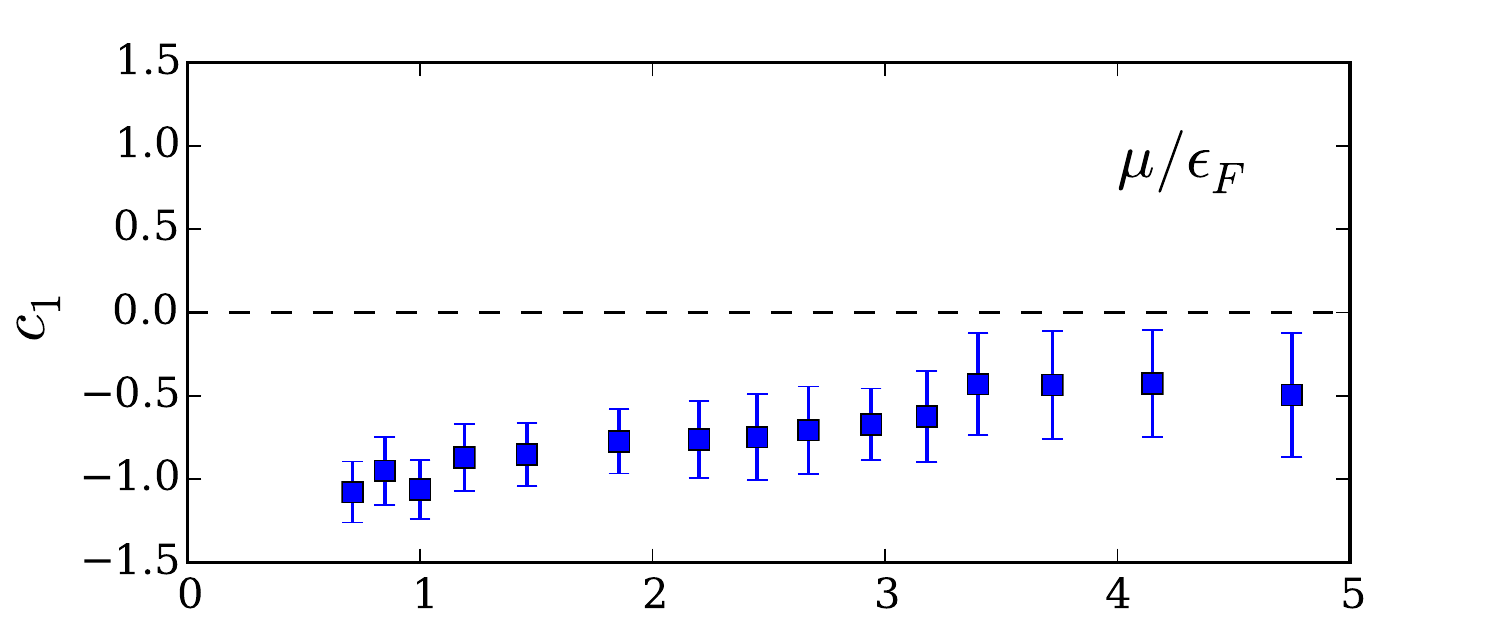}
\includegraphics[width=\columnwidth]{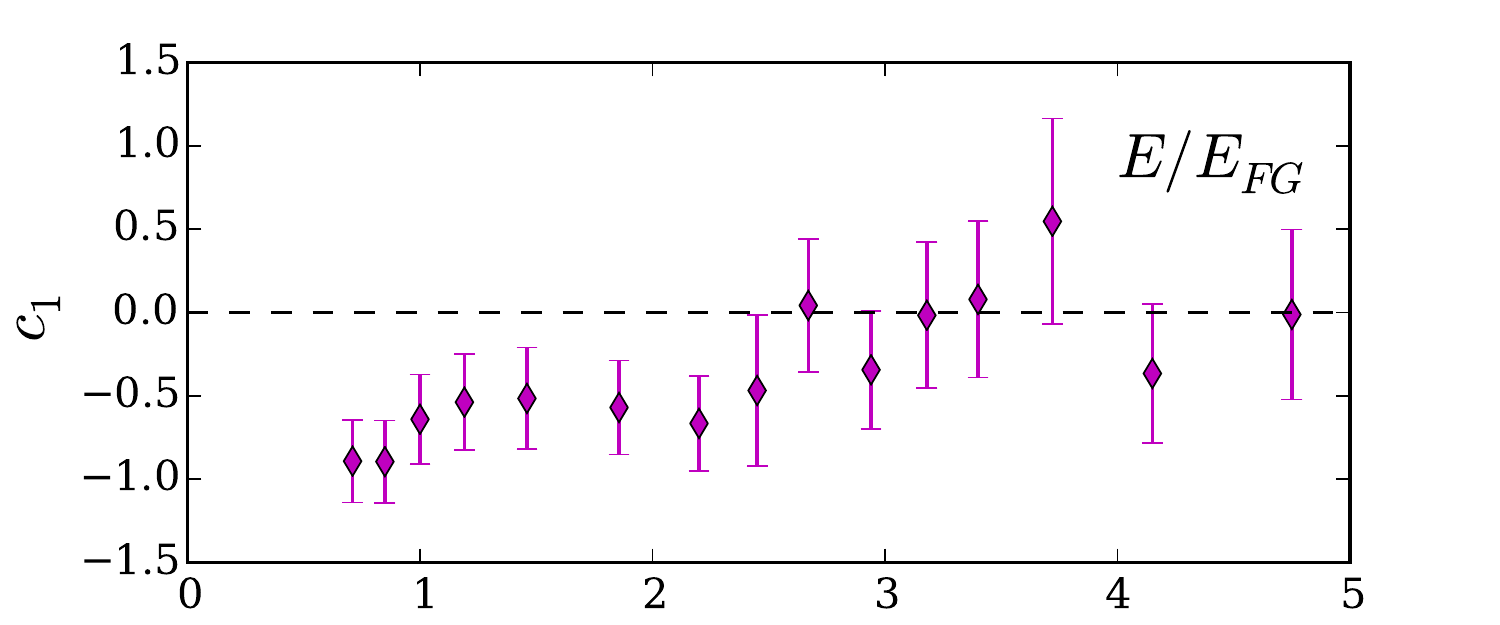}
\includegraphics[width=\columnwidth]{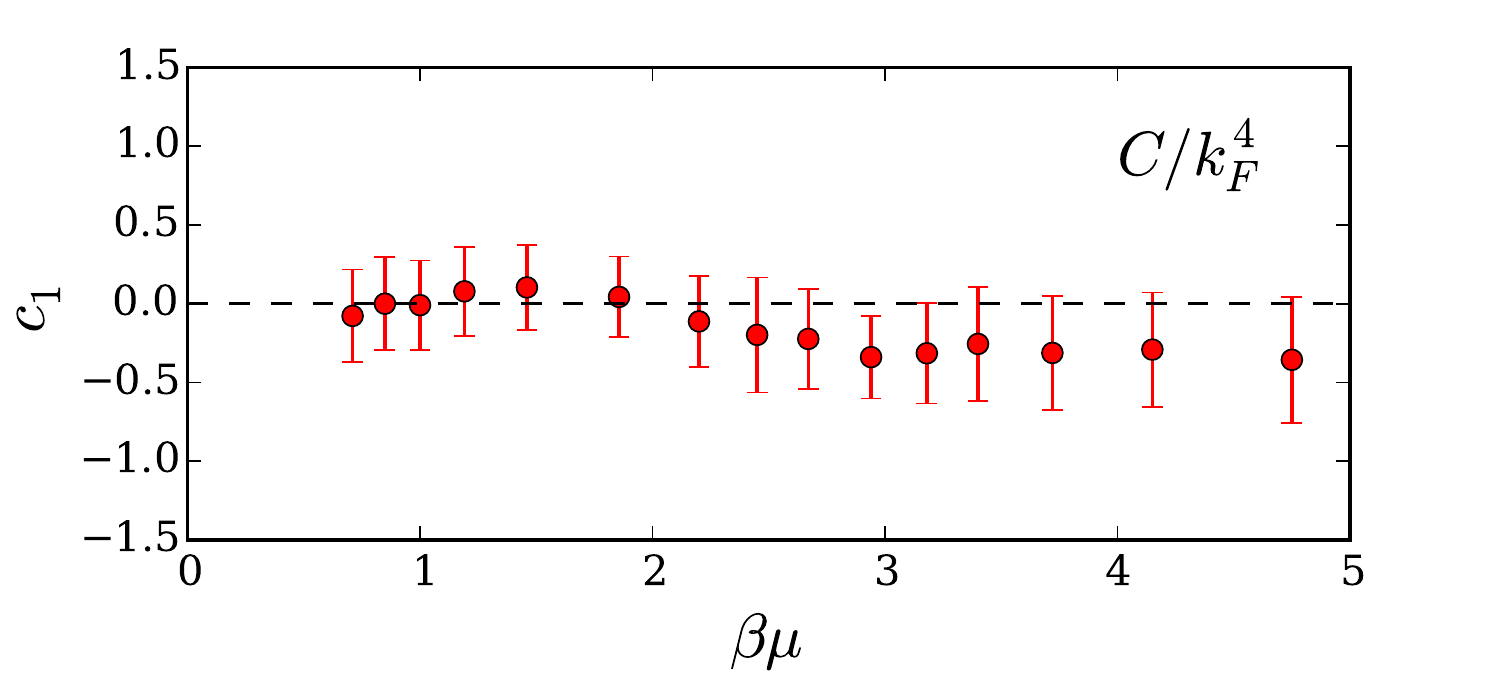}
\caption{\label{fig:linearcoeff}The fit parameter $c_1$ obtained from
  continuum extrapolations, see Eq.~(\ref{eq:continuum}), of Monte
  Carlo data for the chemical potential (top), energy per particle
  (middle), and contact density (bottom). }
\end{figure}

Summarizing the size of discretization error for the continuum extrapolations, we show the values obtained for the coefficient
$c_1$ of $\nu^{1/3}$ in Eq.~(\ref{eq:continuum}).  Figure~\ref{fig:linearcoeff}
shows that the Monte Carlo data for the chemical potential and energy per
particle have significant linear dependence on the lattice spacing, as
parametrized by $\nu^{1/3}$, especially at lower $\beta\mu$.  This is similar
to what was seen for lattice determinations of $T_c$ \cite{burovski,ourmain}.
Nevertheless, the fits described above, which include possible contributions
of higher-order terms, include estimates of these lattice artifacts in the
uncertainties quoted below. We tried a variety of other fits, altering the
bands in $\beta\mu$, and omitting data with large filling factor,
e.g., $\nu^{1/3} > 0.5$ \cite{fermihubbardcontlim2}. These variations produced fits which were in agreement with our final results, but were less
precise in some cases.

\section{Results}

\subsection{Chemical potential}
\begin{figure*}
\includegraphics[width=\columnwidth]{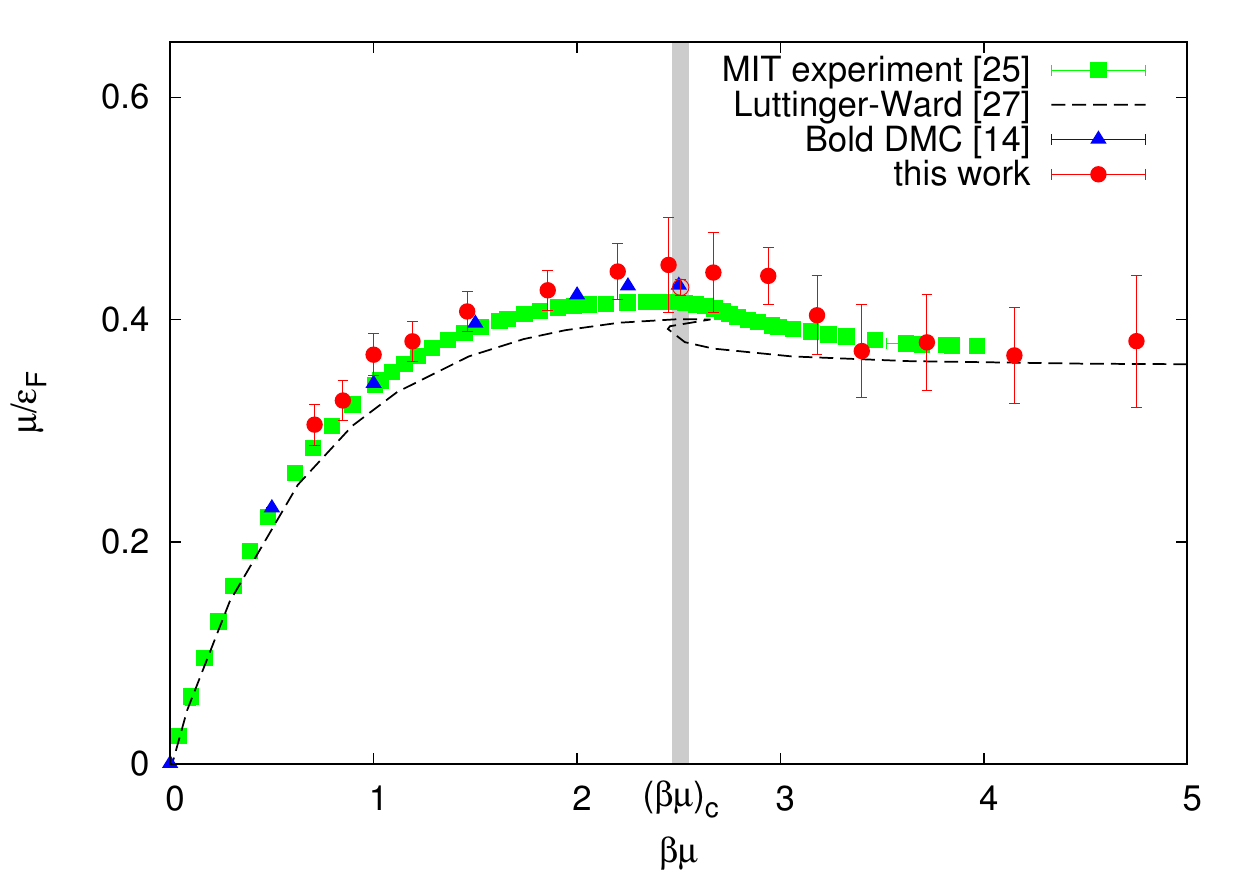}
\includegraphics[width=\columnwidth]{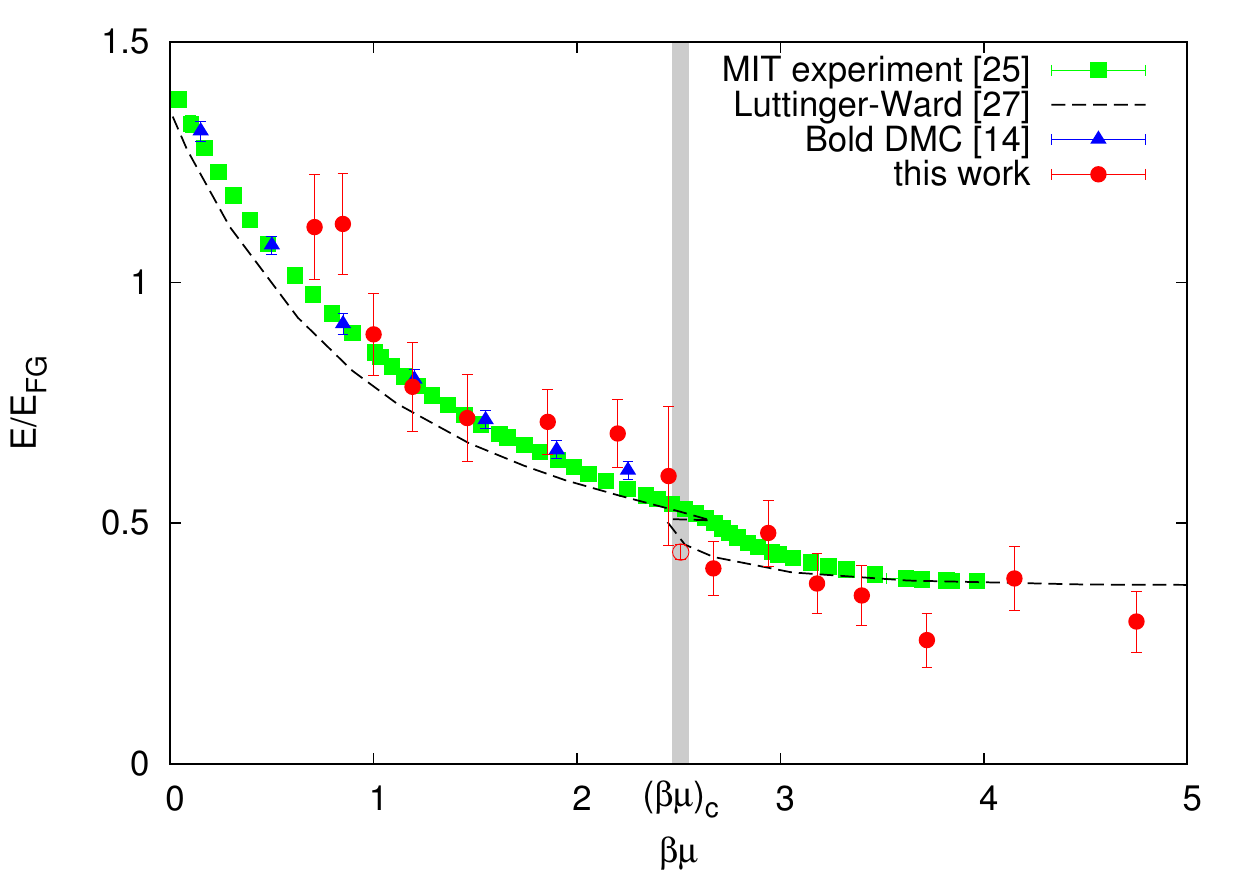}
\caption{\label{fig:chempoten}The chemical potential $\mu/\ef$ (left panel) and the energy per particle $E/E_{\rm FG}$ (right panel) in the continuum limit versus $\beta\mu$. We compare our results (red circles; the empty circles denote our results at $T_c$ from \cite{ourmain}) with experimental data \cite{zwierleinTc} (green squares), as well as results obtained with bold DMC \cite{vanhouckewerneretal} (blue triangles) and with the Luttinger-Ward formalism \cite{selfconsistent} (black dashed line). The gray bar indicates the critical point \cite{ourmain} with error margin.}
\end{figure*}

The left panel of Fig.~\ref{fig:chempoten} shows the continuum limit of the chemical potential as a function of $\beta\mu$. We see excellent agreement with experimental data \cite{zwierleinTc, salomon}, as well as with several other theoretical predictions \cite{vanhouckewerneretal, selfconsistent}. Our results below $T_c$ capture the experimentally observed change of the slope of the chemical potential curve.

\subsection{Energy per particle}
The energy is composed of the kinetic energy $E_{\textnormal{kin}}$ and the
interaction energy $E_{\textnormal{int}}$. We find that, within the statistical uncertainties, neither $E_{\mathrm{kin}}/N$ nor $E_{\mathrm{int}}/N$ exhibit dependence on $L$. This insensitivity can be understood by looking at the lattice Monte Carlo estimator for $E_{\textnormal{kin}}/N=E_{\textnormal{kin}}/L^3\nu$ \cite{ourmain}, which can be expressed as
\begin{equation}
\frac{E_{\textnormal{kin}}}{L^3\nu}=6\left(1-\frac{\sum_\sigma\langle c^\dagger_{\mathbf{x}\sigma}c_{(\mathbf{x}+\hat{\mathbf{j}})\sigma}\rangle}{\nu}\right).
\end{equation}
This expression contains the ratio of the kinetic energy operator $\sum_\sigma\langle c^\dagger_{\mathbf{x}\sigma}c_{(\mathbf{x}+\hat{\mathbf{j}})\sigma}\rangle$ and the filling factor operator $\sum_\sigma\langle  c^\dagger_{\mathbf{x}\sigma}c_{\mathbf{x}\sigma}\rangle$. These two operators have a very similar structure, which explains to some extent why finite-volume errors are negligible within the error bars of the data. The same holds for the interaction part of the energy. Therefore it is sufficient to consider the finite-size scaling of $1/\ef$ (which follows
directly from the finite-size scaling of $\nu$), while the data for $E/L^3\nu$
obtained at different lattice sizes can be fitted by a constant. Our data
confirms this scaling, as the constant fits of $E/L^3\nu$ yield acceptable
$\chi^2$-values, see the middle panel of Figs.~\ref{fig:TDextrapols_lowT} and \ref{fig:TDextrapols_highT} for several examples.

The results for the energy per particle $E/E_{\rm FG}$, where $E_{\rm FG}=(3/5)N\ef$ is the ground state energy of the free gas, are shown in the right panel of Fig.~\ref{fig:chempoten}. Like for the chemical potential, we obtain excellent agreement with experimental data \cite{zwierleinTc, salomon} and theory \cite{vanhouckewerneretal, selfconsistent}.

\subsection{Contact density}
The contact density can be interpreted as a measure of the local pair density \cite{braaten}. The contact plays a crucial role for several universal relations derived by Tan \cite{tan1, tan2, tan3}. We use the definition $C=m^2g_0E_{\textnormal{int}}$, where $g_0$ is the physical coupling constant \cite{braaten, wernercastin3}. The contact is related to the contact density $\mathcal{C}$ via $C=\int\mathcal{C}(\mathbf{r})d^3r$, or for homogeneous systems simply $C=\mathcal{C}V$.

In \cite{ourproc2} we have presented results for the contact density at the critical point. Now we extend this study to other values of the temperature. For the finite-size scaling we can rewrite the dimensionless contact density as
\begin{equation}
\frac{\mathcal{C}}{\ef^2}=\frac{UE_{\textnormal{int}}}{4L^3\ef^2}=\frac{U}{4}\,\frac{E_{\mathrm{int}}}{N} \, \frac{\nu^{-1/3}}{(3\pi^2)^{4/3}}\propto\nu^{-1/3}\frac{E_{\textnormal{int}}}{N}.
\end{equation}
Since $E_{\rm int}/N$ is independent of $L$ within uncertainties, this part
of the contact density for different lattice sizes is fit to a
constant (see the right panel of Figs.~\ref{fig:TDextrapols_lowT} and \ref{fig:TDextrapols_highT} for examples of such fits), while the thermodynamic limit for the part proportional to $\nu^{-1/3}$ follows from the finite-size scaling of the filling factor $\nu$.
\begin{figure}
\begin{center}
\includegraphics[width=\columnwidth]{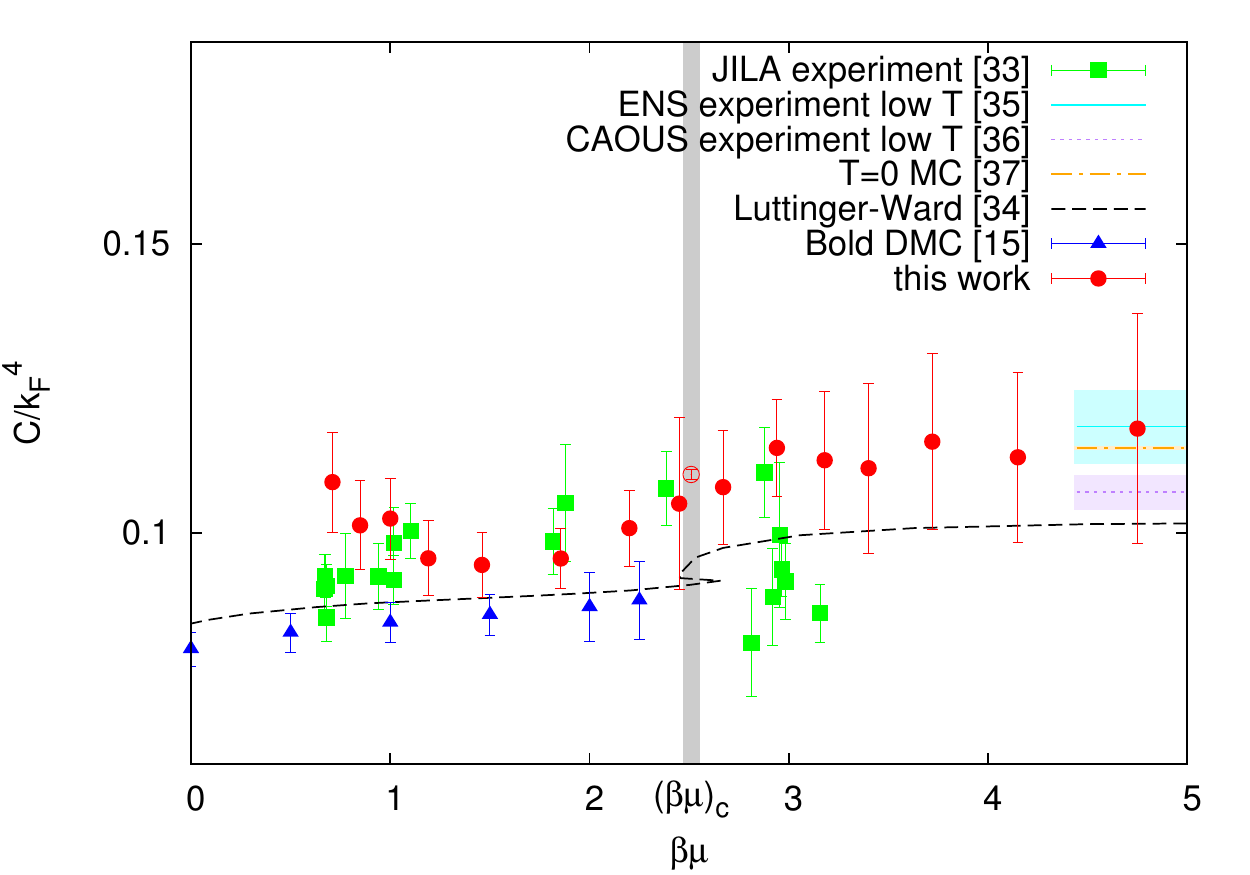}
\end{center}
\caption{\label{fig:CvsT}The contact density $\mathcal{C}/\ef^2=\mathcal{C}/\kf^4$ in the continuum limit versus $\beta\mu$. We compare our results (red circles; the empty circle denotes our result at $T_c$ from \cite{ourproc2}) with experimental data \cite{Sagi:2012jin} (green squares), as well as results obtained with bold DMC \cite{BDMCvanhoucke2013} (blue triangles), Luttinger-Ward formalism \cite{ensscontact} (black dashed line) and the zero-temperature results from \cite{salomoncontact} (cyan line with error margin), \cite{Hoinka:2012aaa} (purple dotted line with error margin) and \cite{gandolficontact} (orange dash-dotted line with error margin).}
\end{figure}

Figure~\ref{fig:CvsT} shows the contact density in the continuum limit. There has been recent progress experimentally investigating Tan's contact, mostly for trapped systems \cite{Hoinka:2012aaa,kuhnle2011,stewart2010} as well as numerical and analytical calculations \cite{BDMCvanhoucke2013,Boettcher2013,ensscontact,hucontact,tmatrixcontact,contactdrut}. The homogeneous contact in the normal phase has been studied experimentally in Ref.~\cite{Sagi:2012jin}. They find a sharp decrease in the contact around the superfluid phase transition. We do not observe any such sudden change around $T_c$, but our results above $T_c$ show good agreement with their data. Our results at low temperature also show excellent agreement with the zero-temperature numerical \cite{gandolficontact} and experimental results \cite{Hoinka:2012aaa,salomoncontact} (the contact is not discussed explicitly in the latter reference, but can be easily extracted with the appropriate Tan relation yielding $\mathcal{C}/\ef ^ 2 = 2\zeta/5\pi=0.1184(64)$).

\section{Summary and Conclusion}
In summary, we have calculated the chemical potential, the energy density and the contact of a homogeneous 3d balanced unitary Fermi gas at different temperatures above and below the critical point. Our results show good agreement with experimental measurements and provide a benchmark for future studies, in particular below $T_c$ where few accurate predictions and measurements are available.

\begin{acknowledgments}
We thank Evgeni Burovski, Tilman Enss, Mark Ku, Rabin Paudel,
Nikolay Prokof'ev, Yoav Sagi, Boris Svistunov, Kris Van Houcke, Felix Werner and Martin
Zwierlein for providing us with their data and for helpful discussions. OG
acknowledges support from the NSF under Grant No. PHY-1314735. MW is
supported by the Science and Technologies Facilities Council.
\end{acknowledgments}

\bibliography{allbib}

\end{document}